\documentclass[%
 reprint,
superscriptaddress,
preprintnumbers,
nofootinbib,
 amsmath,amssymb,
]{revtex4-2}

\usepackage{xcolor}% Color text
\usepackage{graphicx}% Include figure files
\newcommand{\planck}[0]{\textit{Planck}}

\newcommand{\npipe}[0]{\texttt{NPIPE}}

\begin{document}

\preprint{RESCEU-11/23}

\title{Constraint on Early Dark Energy from Isotropic Cosmic Birefringence}

\author{Johannes R. Eskilt}
\email{j.r.eskilt@astro.uio.no}
\affiliation{Institute of Theoretical Astrophysics, University of Oslo, P.O. Box 1029 Blindern, N-0315 Oslo, Norway}
\affiliation{Imperial Centre for Inference and Cosmology, Department of Physics, Imperial College London, Blackett Laboratory, Prince Consort Road, London SW7 2AZ, United Kingdom}
\author{Laura Herold}
\affiliation{
Max Planck Institute for Astrophysics, Karl-Schwarzschild-Str. 1, D-85748 Garching, Germany
}
\author{Eiichiro Komatsu}
\affiliation{
Max Planck Institute for Astrophysics, Karl-Schwarzschild-Str. 1, D-85748 Garching, Germany
}
\affiliation{Kavli IPMU (WPI), UTIAS, The University of Tokyo, Kashiwa, 277-8583, Japan}
\author{Kai Murai}
\affiliation{ICRR, The University of Tokyo, Kashiwa, 277-8582, Japan}
\affiliation{Kavli IPMU (WPI), UTIAS, The University of Tokyo, Kashiwa, 277-8583, Japan}
\author{Toshiya Namikawa}
\affiliation{Kavli IPMU (WPI), UTIAS, The University of Tokyo, Kashiwa, 277-8583, Japan}
\author{Fumihiro Naokawa}
\affiliation{Department of Physics, Graduate School of Science, The University of Tokyo, Bunkyo-ku, Tokyo 113-0033, Japan}
\affiliation{Research Center for the Early Universe, The University of Tokyo, Bunkyo-ku, Tokyo 113-0033, Japan}

\date{\today}% It is always \today, today,
\begin{abstract}
Polarization of the cosmic microwave background (CMB) is sensitive to new physics violating parity symmetry, such as the presence of a pseudoscalar ``axionlike'' field. Such a field may be responsible for early dark energy (EDE), which is active prior to recombination and provides a solution to the so-called Hubble tension. The EDE field coupled to photons in a parity-violating manner would rotate the plane of linear polarization of the CMB and produce a cross-correlation power spectrum of $E$- and $B$-mode polarization fields with opposite parities. In this paper, we fit the $EB$ power spectrum predicted by the photon-axion coupling of the EDE model with a potential $V(\phi)\propto [1-\cos(\phi/f)]^3$ to polarization data from \textit{Planck}. We find that the unique shape of the predicted $EB$ power spectrum is not favored by the data and obtain a first constraint on the photon-axion coupling constant, $g=(0.04\pm 0.16)M_{\text{Pl}}^{-1}$ (68\%~CL),  for the EDE model that best fits the CMB and galaxy clustering data. 
This constraint is independent of the miscalibration of polarization angles of the instrument or the polarized Galactic foreground emission. Our limit on $g$ may have important implications for embedding EDE in fundamental physics, such as string theory.

\end{abstract}

\maketitle
%%%%%%%%%%%%%%%%%%%%%%%%%%%%%%%%%%%%%%%
\section{\label{sec:intro}Introduction}
%%%%%%%%%%%%%%%%%%%%%%%%%%%%%%%%%%%%%%%
The standard cosmological model, called $\Lambda$CDM, includes new physics beyond the standard model of elementary particles and fields, such as dark matter and dark energy~\cite{weinberg:2008}. Clues to their physical nature may be found in possible deviations from the $\Lambda$CDM model. In recent years, a growing number of such deviations, or ``tensions,'' have been reported~\cite{abdalla/etal:2022}, which may point toward new physics. In this paper, we study a fascinating connection between two hints of new physics: early dark energy (EDE) as a solution to the so-called Hubble tension (see Refs.~\cite{kamionkowski/riess:prep,poulin/etal:prep} for reviews) and cosmic birefringence, a rotation of the plane of linear polarization of photons (see Ref.~\cite{komatsu:2022} for a review).

EDE, which was active prior to the epoch of recombination at a redshift of $z\simeq 1090$~\cite{doran/etal:2001,wetterich:2004,doran/robbers:2006}, can resolve the Hubble tension~\cite{karwal/kamionkowski:2016,poulin/etal:2019} by modifying the value of the Hubble constant, $H_0$, inferred from the cosmic microwave background (CMB) data and brings it into agreement with $H_0$ inferred from the local distance ladder~\cite{riess/etal:2022}. But is EDE \textit{the} solution to the Hubble tension? To make progress, one must look elsewhere for corroborating evidence.

In this paper, we search for a signature of EDE in the polarization of the CMB. If the EDE field, $\phi$, is a pseudoscalar ``axionlike'' field, it could couple to electromagnetism in a parity-violating manner in the Lagrangian density, ${\mathcal L}$. We write~\cite{ni:1977,turner/widrow:1988}
\begin{equation}
\label{eq:lagrangian}
    {\mathcal L}=-\frac12(\partial\phi)^2-V(\phi)-\frac14F_{\mu\nu}F^{\mu\nu}-\frac14 g\phi F_{\mu\nu}\tilde F^{\mu\nu}\,,
\end{equation}
where $g$ is the photon-axion coupling constant, and  $F_{\mu\nu}$ and $\tilde F^{\mu\nu}$ are the field strength tensor of the photon field and its dual tensor, respectively. 

The last term in Eq.~\eqref{eq:lagrangian} is a Chern-Simons term, which violates parity symmetry in the presence of a spacetime-dependent condensate of $\phi$.
This term appears naturally for an axionlike field with $g=c_{\phi\gamma}\alpha_\text{em}/(2\pi f)$, where $c_{\phi\gamma}$ is an anomaly coefficient, $\alpha_\text{em}\simeq 1/137$ the electromagnetic fine-structure constant, and $f$  the axion decay constant (see, e.g., Eq.~(24) of Ref.~\cite{marsh:2016}), and has been considered for EDE models in Refs.~\cite{capparelli/caldwell/melchiorri:2020,fujita/etal:2021b,Nakagawa:2022knn,Murai:2022zur,Greco:2022xwj,galaverni/finelli/paoletti:2023}. 

We assume a ``canonical'' EDE potential,
${V(\phi)=V_0[1-\cos(\phi/f)]^3}$, where $V_0$ is the normalization. This model can resolve the Hubble tension~\cite{poulin/etal:2019,smith/poulin/amin:2020,smith/eal:2021,murgia/abellan/poulin:2021,smith/etal:2022,herold/ferreira/komatsu:2022,simon/etal:2023,herold/ferreira:2023}. See Refs.~\cite{hill/etal:2020,ivanov/etal:2020,damico/etal:2021,hill/etal:2022, LaPosta:2021pgm, reeves/etal:2023, Cruz:2023cxy,goldstein/etal:prep} for other constraints on this model. For other EDE models that can resolve the Hubble tension, see Ref.~\cite{poulin/etal:prep} and references therein.

To probe violation of parity symmetry in the polarization pattern of the CMB, one can decompose a pixelized map of the observed Stokes parameters into eigenstates of parity called $E$ and $B$ modes~\cite{zaldarriaga/seljak:1997,kamionkowski/etal:1997}:
\begin{equation}
    Q(\hat{\mathbf n})\pm iU(\hat{\mathbf n})=-\sum_{\ell=2}^{\ell_\text{max}}\sum_{m=-\ell}^\ell\left(E_{\ell m}\pm iB_{\ell m}\right){}_{\pm 2}Y_\ell^m(\hat{\mathbf n})\,,
\end{equation}
where $\hat{\mathbf n}$ is the direction of an observer's line of sight, $E_{\ell m}$ and $B_{\ell m}$ are the spherical harmonics coefficients of the $E$ and $B$ modes, respectively, ${}_{\pm 2}Y_\ell^m(\hat{\mathbf n})$ are the spin-2 spherical harmonics, and $\ell_\text{max}$ is the maximum multipole used for the analysis. The coefficients transform under inversion of spatial coordinates, $\hat{\mathbf n}\to -\hat{\mathbf n}$, as $E_{\ell m}\to (-1)^\ell E_{\ell m}$ and $B_{\ell m}\to (-1)^{\ell+1} B_{\ell m}$. The cross-power spectrum of the $E$ and $B$ modes, $C_\ell^{EB}\equiv (2\ell+1)^{-1}\sum_{m}\operatorname{Re}(E_{\ell m}B_{\ell m}^*)$, has odd parity and is sensitive to parity violation~\cite{lue/wang/kamionkowski:1999}. 

In this paper, we use the $EB$ power spectrum presented in Ref.~\cite{Eskilt:2022cff} to constrain the Chern-Simons term in Eq.~\eqref{eq:lagrangian}. Specifically, this term makes the phase velocities of the right- and left-handed circular polarization states different, which rotates the plane of linear polarization by an angle ${\beta(\hat{\mathbf n})=\frac12g\left[\phi(\eta_\text{o})-\phi(\eta_\text{e},r\hat{\mathbf n})\right]}$ as the photons have traveled from the conformal time of emission, $\eta_\text{e}$, to the observation, $\eta_\text{o}$~\cite{carroll/field/jackiw:1990,carroll/field:1991,harari/sikivie:1992}. Here, $r=c(\eta_\text{o}-\eta_\text{e})$ is the conformal distance to the emitter. As there is no evidence for anisotropic birefringence~\cite{contreras/boubel/scott:2017,namikawa/etal:2020,bianchini/etal:2020,gruppuso/etal:2020,bortolami/etal:2022}, we assume that $\beta$ is independent of $\hat{\mathbf n}$. See Refs.~\cite{capparelli/caldwell/melchiorri:2020,Greco:2022xwj} for a study on anisotropic birefringence from the EDE field.

A distinct feature of the EDE field is that $\phi$ evolves significantly during the epoch of recombination, which has profound implications for observational tests of EDE models with photon-axion coupling~\cite{Nakatsuka:2022epj,Murai:2022zur,galaverni/finelli/paoletti:2023}. If $\phi$ were constant during the epoch of recombination and evolved only later, the observed $E$ and $B$ modes (denoted by the superscript ``o'') would be given by 
$E_{\ell m}^{\textrm{o}}=E_{\ell m}\cos(2\beta)-B_{\ell m}\sin(2\beta)$ and $B_{\ell m}^{\textrm{o}}=E_{\ell m}\sin(2\beta)+B_{\ell m}\cos(2\beta)$, respectively, and~\cite{feng/etal:2005}
\begin{equation}
    \label{eq:simple_eb}
    C_\ell^{EB, {\textrm{o}}} = \frac{\sin(4\beta)}{2}\left( C^{EE}_\ell - C^{BB}_\ell\right)+\cos(4\beta)C_\ell^{EB}\,,
\end{equation}
where the last term is the intrinsic $EB$ correlation at the time of emission, and $C_\ell^{EE}$ and $C_\ell^{BB}$ are the auto-power spectra of $E$ and $B$ modes at emission, respectively. 

Eq.~\eqref{eq:simple_eb} has been assumed in all of the previous constraints on isotropic cosmic birefringence, including Refs.~\cite{minami/komatsu:2020b,NPIPE:2022,Eskilt:2022wav,Eskilt:2022cff}, which reported a tantalizing hint of $\beta$ with a statistical significance of $>3\sigma$. However, if $\phi$ evolved significantly during the epoch of recombination, $C^{EB, \mathrm{o}}_{\ell}$ would no longer be given by Eq.~\eqref{eq:simple_eb} and would exhibit complex dependence on $\ell$~\cite{liu/lee/ng:2006,finelli/galaverni:2009}, allowing us to distinguish between different origins of cosmic birefringence~\cite{Nakatsuka:2022epj,Murai:2022zur,galaverni/finelli/paoletti:2023}. 

In this paper, we present a first constraint on the photon-axion coupling constant, $g$, from the \textit{shape} of the $EB$ power spectrum. This is a powerful new approach that breaks the degeneracy between cosmic birefringence and an instrumental miscalibration of polarization angles of detectors, $\alpha$. Specifically, Eq.~\eqref{eq:simple_eb} can be generated not only by cosmic birefringence, but also by $\alpha$~\cite{wu/etal:2009,miller:2009,WMAP:2011,RAC:2022}. Therefore, 
$\beta$ in Eq.~\eqref{eq:simple_eb} needs to be replaced by the sum, $\alpha+\beta$, and we cannot distinguish between $\alpha$ and $\beta$ unless one calibrates $\alpha$ well~\cite{Cornelison:2022zrc} or uses other information such as the Galactic foreground emission~\cite{minami/etal:2019} and the so-called ``reionization bump'' at $\ell\lesssim 10$~\cite{sherwin/namikawa:2023}.
Due to this complication, the current hint of $\beta$ remains somewhat inconclusive~\cite{diego-palazuelos/etal:2023}. Our new approach, based on the shape of the $EB$ power spectrum different from $C_\ell^{EE}-C_\ell^{BB}$, is free from this complication. 

%%%%%%%%%%%%%%%%%%%%%%%%%%%%%%%%%%%%
\section{\label{sec:pipeline}Method}
%%%%%%%%%%%%%%%%%%%%%%%%%%%%%%%%%%%%
We use the official Public Release 4 (often called ``\npipe-processed'' data) full-sky polarization maps from the High-Frequency Instrument (HFI) of the \planck\ mission at
frequencies of $\nu=100$, 143, 217, and 353\,GHz~\cite{Planck2018III, PlanckIntLVII}. The \npipe\ processing improved the sensitivity of frequency maps by including more data and better instrumental modeling compared to previous \planck\ data releases. In addition to releasing full-mission and time-split maps, \npipe\ also divided the set of detectors for each frequency band into two groups, and produced 8 full-sky maps in total. We work with these detector-split maps as their cross-correlations yield less correlated noise and instrumental systematics than time-split maps.

Unlike in Ref.~\cite{Eskilt:2022cff}, we do not include WMAP or \planck\ Low-Frequency Instrument data for simplicity of the data analysis. Including them would modestly improve the constraints by about 10\%.

We use two masks~\cite{Eskilt:2022wav, Eskilt:2022cff}. One is a small mask, leaving nearly full-sky data available for the analysis. The other is a large mask, removing 30\% of the Galactic plane. Both masks remove pixels containing polarized point sources given in official \planck point-source maps and pixels where the carbon-monoxide (CO) emission is brighter than $45\,\mathrm{ K_{RJ}\, km\, s^{-1}}$. These masks leave the sky fractions of $f_\text{sky} = 0.92$ and 0.62 for small and large masks available for analysis, respectively.

We calculate the $EB$ power spectra of 8 masked polarization maps using \texttt{PolSpice}\footnote{\url{http://www2.iap.fr/users/hivon/software/PolSpice/}} \citep{Chon:2003gx}. We then beam-deconvolve the $EB$ power spectra using the official \npipe\ beam transfer functions and the pixel window functions from the \texttt{HEALPix} library~\cite{gorski/etal:2005}.

We calculate the inverse-variance weighted average of the beam-deconvolved $EB$ power spectra from
\begin{equation}
    \bar{C}^{EB, \textrm{o}}_{b} \equiv \textrm{E}\left(C^{EB, \textrm{o}}_b \right) = \frac{\vec{1} \cdot \textbf{M}^{-1}_b \cdot \vec{C}^{EB}_b}{\vec{1} \cdot \textbf{M}^{-1}_b \cdot \vec{1}}\,,
    \label{eq:stackedEB}
\end{equation}
which we call a ``stacked $EB$ power spectrum''~\cite{Eskilt:2022cff}. The variance is given by
\begin{equation}
    \textrm{Var}\left(C^{EB, \textrm{o}}_b \right) = \frac{1}{\vec{1} \cdot \textbf{M}^{-1}_b \cdot \vec{1}}\,.
\end{equation}
Here, $\vec{1}$ is a unit vector, $\vec{C}^{EB}_b$ is the binned set of all combinations of the observed beam-deconvolved $EB$ power spectra, $C^{E_i B_j, \mathrm{o}}_{\ell}$, where $i$ and $j$ denote frequency bands, and $\textbf{M}_b$ is the binned covariance matrix for $\vec{C}^{EB}_b$. 

We bin $C^{E_i B_j, \mathrm{o}}_{\ell}$ and $\textbf{M}_{\ell}$ as
\begin{equation}
    \vec{C}^{EB}_b = \frac{1}{\Delta \ell} \sum_{\ell\in b} \vec{C}^{EB}_\ell\,, \quad
    \textbf{M}_b = \frac{1}{\Delta \ell^2} \sum_{\ell\in b} \textbf{M}_\ell\,,
\end{equation}
where
\begin{equation}
    \textbf{M}_{\ell} = \textrm{Cov}\left(C^{E_i B_j, \mathrm{o}}_{\ell}, C^{E_p B_q, \mathrm{o}}_{\ell} \right)
    \simeq \frac{C^{E_i E_p, \mathrm{o}}_{\ell} C^{B_j B_q, \mathrm{o}}_{\ell} }{(2\ell + 1)f_{\rm sky}}\,.
\end{equation}
We neglected the term $C^{E_i B_q, \mathrm{o}}_{\ell}\ C^{E_p B_j, \mathrm{o}}_{\ell}$, which is much smaller than the other term. As we use the observed power spectra in $\textbf{M}_{\ell}$, this term fluctuates around zero and biases $\textbf{M}_{\ell}$ when included; thus, it is best to neglect it~\cite{minami/etal:2019}.
We also excluded auto-power spectra ($i = j$) due to noise domination at high $\ell$ for the $EE$ and $BB$ power spectra. This does not affect $C^{E_i B_j}_{b}$ directly, but it does affect $\textbf{M}_{b}$. Following the previous work~\cite{PlanckIntXLIX,minami/komatsu:2020b,NPIPE:2022,Eskilt:2022wav,Eskilt:2022cff}, we bin the power spectra over 20 multipoles, i.e., $\Delta \ell = 20$. Our multipole range is from $\ell_{\text{min}}=51$ to $\ell_{\text{max}}=1490$, which gives a total of 72 bins. The stacked $EB$ power spectrum is publicly available\footnote{\url{https://github.com/LilleJohs/Observed-EB-Power-Spectrum/}}.

The aim of this paper is to constrain the coupling, $g$, to pre-recombination EDE while marginalizing over post-recombination cosmic birefringence and miscalibration angles, $\alpha+\beta$. To this end, we fit the stacked $EB$ power spectrum for $g$ and $\alpha + \beta$ simultaneously. We sample these parameters using a Markov Chain Monte Carlo sampler \texttt{emcee} \citep{ForemanMackey:2012ig}. The log-likelihood function is $-2\ln \mathcal{L} = \sum_{b} v^2_{b} / \text{Var}\left(C^{EB, \textrm{o}}_b \right) $, where
\begin{eqnarray}
    \nonumber
    v_b &\equiv &\bar{C}^{EB, o}_{b} - \cos\left[4(\alpha+\beta)\right]gM_\text{Pl} C_b^{EB, \text{EDE}} \\
    &-& \frac{\sin{\left[4(\alpha+\beta)\right]}}{2}\left(C^{EE, \textrm{CMB}}_b-C^{BB, \textrm{CMB}}_b \right)\,.
    \label{eq:EBmodel}
\end{eqnarray}
We compute the CMB $EE$ and $BB$ power spectra using \texttt{CAMB}\footnote{\url{https://github.com/cmbant/CAMB}} \citep{Lewis:2000} with the best-fitting $\Lambda$CDM parameters of the \planck\ 2018 analysis \cite{Planck2018VI}. The EDE $EB$ power spectrum is computed using a modified version of the CLASS code~\cite{Blas:2011rf} developed in Refs.~\cite{Murai:2022zur,Naokawa:2023upt} with $g=M_\text{Pl}^{-1}$, as $g$ just gives the amplitude of $C_\ell^{EB, \text{EDE}}$ and can be rescaled later. Here, $M_\text{Pl}\simeq 2.4\times 10^{18}$~GeV is the reduced Planck mass. 

The EDE model has 3 parameters in addition to the standard $\Lambda$CDM parameters: $f_\text{EDE}$, $z_\text{c}$ and $\theta_i$. Here, $f_\text{EDE}$ is the maximum energy density fraction of the EDE field reached at a redshift $z_\text{c}$, while $\theta_i$ is a dimensionless initial value of the EDE field, $\theta_i\equiv \phi_i/f$. The first two are related to the fundamental parameters in the potential, $V_0$ and $f$. 

We fix all EDE and $\Lambda$CDM parameters to the best-fitting parameters from Ref.~\cite{herold/ferreira:2023} for two different data sets (see Table~\ref{table:best-fit}) and sample only $g$ and $\alpha+\beta$. As described in Ref.~\cite{herold/ferreira:2023}, the baseline data set includes the \planck\ temperature and polarization power spectra \cite{Planck2018V} and the galaxy power spectra of the Baryon Oscillation Spectroscopic Survey (BOSS) Data Release 12 \cite{BOSS:2016wmc} (second column), while the second data set additionally includes the SH0ES measurement of $H_0$~\cite{riess/etal:2022} from the local distance ladder method (third column).

\begin{table}
\centering
\begin{tabular}{ c|c c c }
\hline
 & Base & Base+SH0ES \\
\hline
$f_{\text{EDE}}$ & 0.0872 & 0.1271 \\ 
$\log_{10} z_c $& 3.560 & 3.563 \\ 
$\theta_i$& 2.749 & 2.768 \\ 
\hline
$100 \, \omega_{\text{b}}$ & 2.265 & 2.278 \\ 
$\omega_{\text{CDM}}$ & 0.1282 & 0.1324 \\ 
$100 \, \theta_s$ & 1.041 & 1.041 \\ 
$\ln\left(10^{10} A_s\right)$ & 3.063 & 3.071 \\ 
$n_s$ & 0.983 & 0.992 \\ 
$\tau$ & 0.0562 & 0.0568 \\ 
\hline
\end{tabular}
\caption{\label{table:best-fit} Best-fitting cosmological parameters under the \planck\ + BOSS (base) and base + SH0ES data sets~\cite{herold/ferreira:2023}.
}
\end{table}

%%%%%%%%%%%%%%%%%%%%%%%%%%%%%%%%%%%%
\section{\label{sec:results}Results}
%%%%%%%%%%%%%%%%%%%%%%%%%%%%%%%%%%%%

\begin{figure}
\centering
\includegraphics[width=\linewidth]{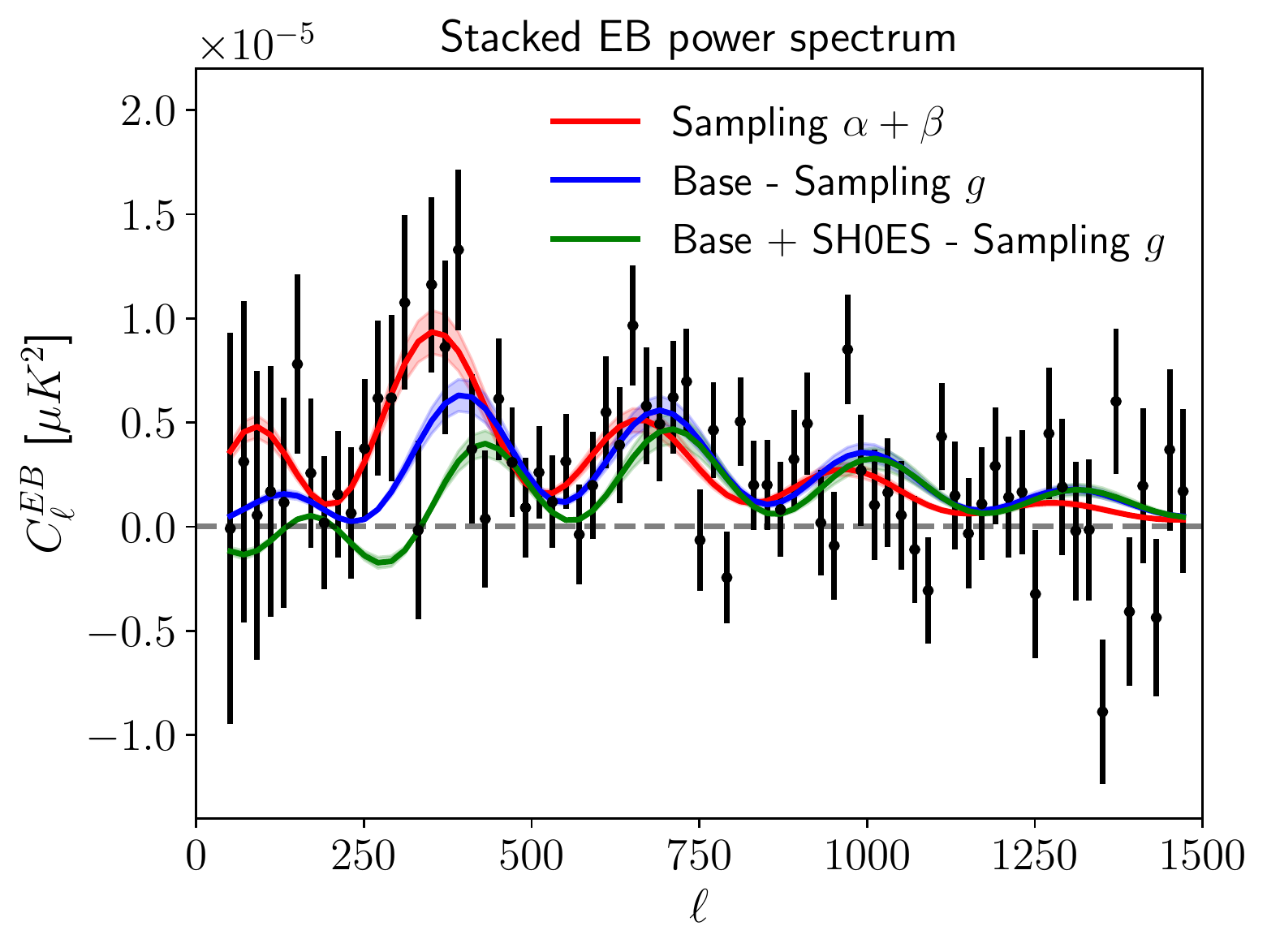}
\caption{\label{fig:g_beta_eb_spectrum} Stacked observed $EB$ power spectrum (black points with error bars), compared to the best-fitting models of $\alpha + \beta$ (red) and $g$ for two EDE models with parameters shown in Table~\ref{table:best-fit} (blue and green). The $\chi^2$ is $65.8$, $77.5$, and $103.5$ for the red, blue, and green lines, respectively, for 71 degrees of freedom.
}
\end{figure}

Before performing a simultaneous inference for $\alpha+\beta$ and $g$, 
we first fit the stacked $EB$ power spectrum for each of these parameters alone. The black points with error bars in Fig.~\ref{fig:g_beta_eb_spectrum} show the stacked $EB$ power spectrum for the nearly full-sky data. The red shaded area shows the $1\sigma$ band from the $\alpha+\beta$ fit, while the blue and green areas show those from $g$ for the two EDE parameter sets. The shapes of the best-fitting $EB$ power spectra are different for all cases, indicating that we can easily distinguish them. 

As shown in Refs.~\cite{Nakatsuka:2022epj,Murai:2022zur,galaverni/finelli/paoletti:2023}, the shape differences in $C_\ell^{E B}$ come from the time evolution of $\phi$. In the EDE cases, $\phi$ flips the sign during the recombination epoch, and photons from the early stage of the recombination make a positive contribution to $C_\ell^{E B}$, while those from the later stage make a negative contribution. Then, $C_\ell^{E B}$ can be negative for some $\ell$ and has peaks at higher $\ell$ compared to the $\alpha + \beta$ fit.
We find that $\alpha+\beta$ fits the data better, especially at the acoustic peak around $\ell \simeq 400$, which is reflected in a lower $\chi^2$ value for this model (see the caption of Fig.~\ref{fig:g_beta_eb_spectrum}). 

\begin{figure}
\centering
\includegraphics[width=\linewidth]{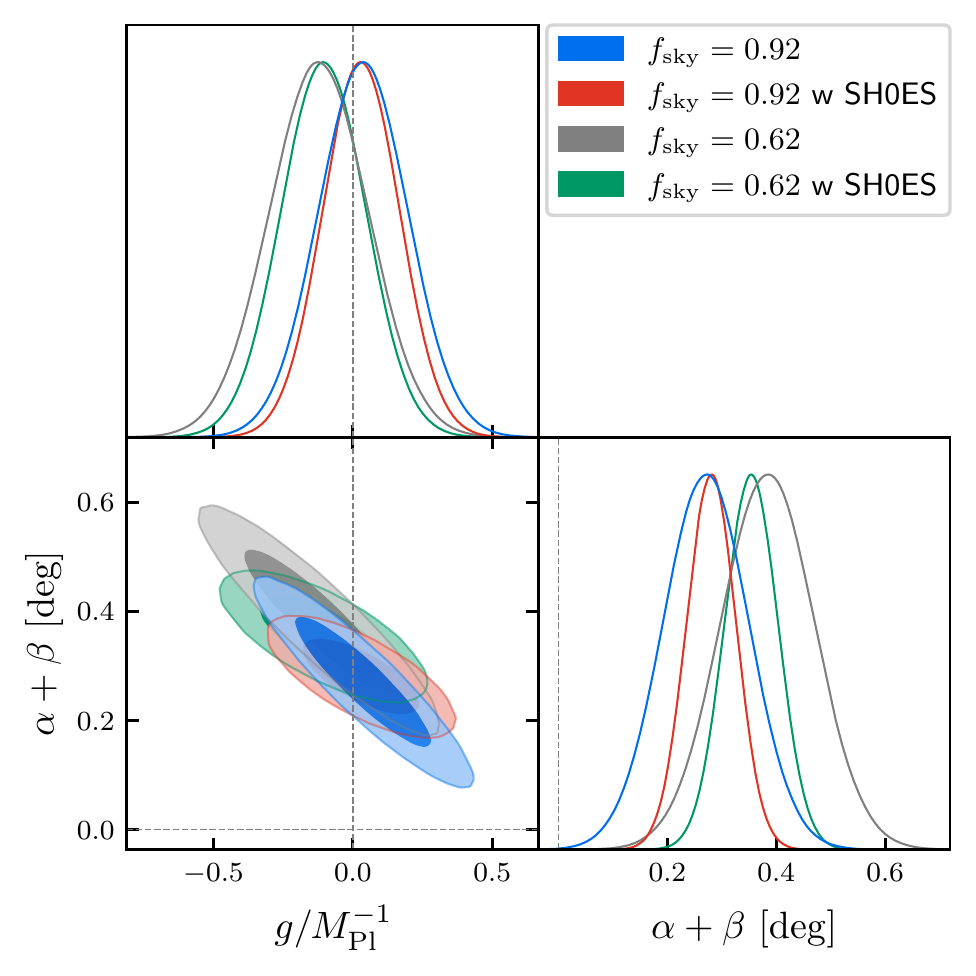}
\caption{\label{fig:g_beta_posterior} Posterior distributions of $g/M_{\text{Pl}}^{-1}$ and $\alpha+\beta$ for the best-fitting EDE parameters under the base and base+SH0ES data sets, and two Galactic masks.
}
\end{figure}

We now present results from jointly sampling $\alpha+\beta$ and $g$. In Fig.~\ref{fig:g_beta_posterior}, we show the posterior distributions of $\alpha+\beta$ and $g$ for the two sets of EDE parameters. 
To show the robustness of the results against the choice of the Galactic mask, the results are shown for both small ($f_{\mathrm{sky}}=0.92$) and large ($f_{\mathrm{sky}}=0.62$) masks. We find that all combinations yield similar results. 

As also examined in Ref.~\cite{Eskilt:2022cff}, we find similar stacked $EB$ power spectra for all sky fractions between $f_{\mathrm{sky}}=0.92$ and 0.62, suggesting that the Galactic foreground emission does not contribute significantly to the stacked $EB$ power spectrum. This is because foreground-dominated channels such as 353~GHz have large $\textbf{M}_b$ and are downweighted in the sum given in Eq.~\eqref{eq:stackedEB}. The foreground $EB$ power spectrum plays a role only when used to calibrate $\alpha$~\cite{NPIPE:2022}. As we do not separate $\alpha$ and $\beta$ in this paper but marginalize over $\alpha+\beta$, our results are insensitive to the Galactic foreground emission.

\begin{figure}
\centering
\includegraphics[width=\linewidth]{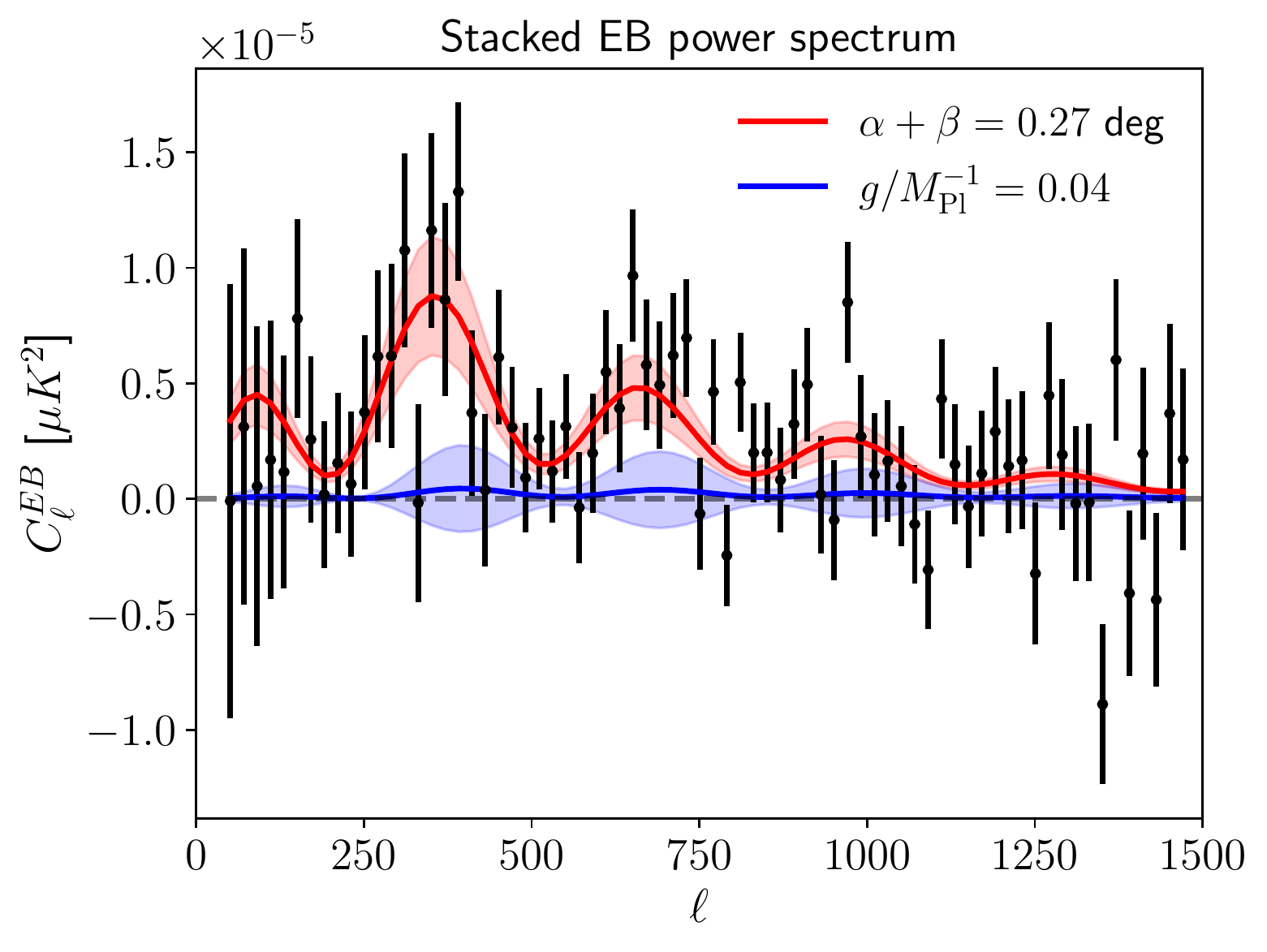}
\caption{\label{fig:best-fitting-g-beta-EB} Quality of the fit for the baseline result. The red area shows the $1\sigma$ band of the $\alpha+\beta$ term, while the blue area shows that of the EDE term given in Eq.~\eqref{eq:EBmodel}. The $\chi^2$ is $65.8$ for 70 degrees of freedom.
}
\end{figure}

We choose the nearly full-sky result with the base EDE parameters as our baseline result, whose quality of the fit is shown in Fig.~\ref{fig:best-fitting-g-beta-EB}. We report $\alpha+\beta = 0.27^\circ \pm 0.08^\circ$ and $g/M_\text{Pl}^{-1} = 0.04 \pm 0.16$ (68\%~C.L.). The former agrees well with those reported in the literature \cite{PlanckIntXLIX,minami/komatsu:2020b,NPIPE:2022,Eskilt:2022wav,Eskilt:2022cff,bortolami/etal:2022}, while the latter is a first constraint on $g$ for the EDE model with $V(\phi)=V_0[1-\cos(\phi/f)]^3$. The data strongly favor $\alpha+\beta$ over $g$ from EDE.

We also check whether the choice of EDE parameters affects the results. We repeat the analysis for the best-fitting cosmological parameters for $f_{\text{EDE}}$ fixed to a grid of values between $0$ to $0.15$ reported in Ref.~\cite{herold/ferreira:2023}. We find that the minimum $\chi^2$ for the stacked $EB$ power spectrum varies less than $\Delta \chi^2 < 0.1$ when sampling $\alpha+\beta$ and $g$ jointly, and obtain similar constraints on the parameters regardless of the value of $f_{\text{EDE}}$. This is because the shape of the observed $EB$ power spectrum is well described by $\alpha+\beta$. This result confirms the robustness of our baseline result.

%%%%%%%%%%%%%%%%%%%%%%%%%%%%%%%%%%%%%%%%%%%
\section{\label{sec:conclusion}Conclusions}
%%%%%%%%%%%%%%%%%%%%%%%%%%%%%%%%%%%%%%%%%%%
We presented a first constraint on the photon-axion coupling constant for the EDE model. Thanks to the unique shape of the predicted $EB$ power spectrum, we were able to derive a constraint on $g/M_\text{Pl}^{-1} = 0.04 \pm 0.16$ (68\%~C.L.) independent of the miscalibration angle or the Galactic foreground emission. 
We find that the \textit{Planck} data do not favor cosmic birefringence caused by a coupling to canonical pre-recombination EDE, $g$, but favor cosmic birefringence that occurred after the epoch of recombination, $\beta$, or a miscalibration of polarization angles of the \textit{Planck} detectors, $\alpha$. 

The derived value of $g$ is much weaker than the strength of the gravitational interaction, $|g|\ll M_\text{Pl}^{-1}$. This may have important implications for embedding EDE in fundamental physics, such as string theory. 
Although previous attempts~\cite{rudelius:2023,mcdonough/scalisi:prep,cicoli/etal:2023} have not yet studied a Chern-Simons coupling for the EDE field in string theory, it is conceivable that such a term appears. 

In particular, Ref.~\cite{cicoli/etal:2023} finds that the best-fitting axion decay constant, $f\simeq 0.2~M_\text{Pl}$, requires fine-tuning of the microscopic parameters to respect the weak gravity conjecture~\cite{Arkani-Hamed:2006emk}, which states that any gauge force must mediate interactions stronger than gravity for some particles~\cite{Harlow:2022gzl}. Similarly, $|g| \ll M_\text{Pl}^{-1}$ may also be in tension with the requirement for a consistent theory of quantum gravity, including the weak gravity conjecture. Whether such a constraint exists on $g$
remains to be studied. See Refs.~\cite{Heidenreich:2021yda,Kaya:2022edp} for some discussion on the weak gravity conjecture in the presence of a Chern-Simons coupling.

The EDE parameters given in Table~\ref{table:best-fit} yield $f = 0.15~M_\text{Pl}$ and $0.18~M_\text{Pl}$ for the base and base+SH0ES parameters, respectively. If we take $g=c_{\phi\gamma}\alpha_\text{em}/(2\pi f)$ with $f = 0.15~M_\text{Pl}$, our constraint on $g$ yields $c_{\phi\gamma}=5.2\pm 21$ (68\%~C.L.). If the weak gravity conjecture demands $|g|\gtrsim M_\text{Pl}^{-1}$, we find $|c_{\phi\gamma}|\gtrsim 130$, which may be too large for an anomaly coefficient, although a natural value for $c_{\phi\gamma}$ depends on the precise mechanism by which the Chern-Simons term arises. In any case, such a large anomaly coefficient is ruled out by our measurement.

Our work opens up a new research area in the $EB$ power spectrum analysis. While we have focused on a particular EDE model with $V(\phi)=V_0[1-\cos(\phi/f)]^n$ for $n=3$, the same analysis can be repeated for any other model, e.g., $n=2$~\cite{galaverni/finelli/paoletti:2023}. As our approach is independent of the miscalibration angle or the Galactic foreground emission, it can be applied to both current and future CMB experiments~\cite{SimonsObservatory:2019,CMB-S4:2019,SPO:2020,NASAPICO:2019,LiteBIRD:2022}.

%%%%%%%%%%%%%%%%%%%%%%%%%%%%%%%%%%%%%%%%%%%
\begin{acknowledgments}
We thank K. Kamada, T. Noumi, I. Obata, M. Scalisi, and M. Shiraishi for useful discussions. This work was supported in part by the European Research Council (ERC) under the Horizon 2020 Research and Innovation Programme (Grant agreement No.~819478), JSPS KAKENHI Grants No.~JP20H05850, No.~JP20H05859, No. JP20J20248 and No. JP22K03682, the Deutsche Forschungsgemeinschaft (DFG, German Research Foundation) under Germany's Excellence Strategy - EXC-2094 - 390783311, the Program of Excellence in Photon Science, and the Forefront Physics and Mathematics Program to Drive Transformation (FoPM), a World-leading Innovative Graduate Study (WINGS) Program, the University of Tokyo. This work has also received funding from the European Union's Horizon 2020 research and innovation programme under the Marie Sk\l odowska-Curie grant agreement No.~101007633. The Kavli IPMU is supported by World Premier International Research Center Initiative (WPI), MEXT, Japan. We acknowledge the use of the Legacy Archive for Microwave Background Data Analysis (LAMBDA), part of the High Energy Astrophysics Science Archive Center (HEASARC). HEASARC/LAMBDA is a service of the Astrophysics Science Division at the NASA Goddard Space Flight Center. \planck\ is a project of the European Space Agency (ESA) with instruments provided by two scientific consortia funded by ESA member states and led by Principal Investigators from France and Italy, telescope reflectors provided through a collaboration between ESA and a scientific consortium led and funded by Denmark, and additional contributions from NASA (USA). Some of the results in this paper have been derived using the \texttt{HEALPix} package~\cite{gorski/etal:2005} and the \texttt{GetDist} package~\cite{Lewis:2019xzd}.
\end{acknowledgments}
\bibliographystyle{apsrev4-2}
\bibliography{references}

\begin{thebibliography}{87}%
\makeatletter
\providecommand \@ifxundefined [1]{%
 \@ifx{#1\undefined}
}%
\providecommand \@ifnum [1]{%
 \ifnum #1\expandafter \@firstoftwo
 \else \expandafter \@secondoftwo
 \fi
}%
\providecommand \@ifx [1]{%
 \ifx #1\expandafter \@firstoftwo
 \else \expandafter \@secondoftwo
 \fi
}%
\providecommand \natexlab [1]{#1}%
\providecommand \enquote  [1]{``#1''}%
\providecommand \bibnamefont  [1]{#1}%
\providecommand \bibfnamefont [1]{#1}%
\providecommand \citenamefont [1]{#1}%
\providecommand \href@noop [0]{\@secondoftwo}%
\providecommand \href [0]{\begingroup \@sanitize@url \@href}%
\providecommand \@href[1]{\@@startlink{#1}\@@href}%
\providecommand \@@href[1]{\endgroup#1\@@endlink}%
\providecommand \@sanitize@url [0]{\catcode `\\12\catcode `\$12\catcode
  `\&12\catcode `\#12\catcode `\^12\catcode `\_12\catcode `\%12\relax}%
\providecommand \@@startlink[1]{}%
\providecommand \@@endlink[0]{}%
\providecommand \url  [0]{\begingroup\@sanitize@url \@url }%
\providecommand \@url [1]{\endgroup\@href {#1}{\urlprefix }}%
\providecommand \urlprefix  [0]{URL }%
\providecommand \Eprint [0]{\href }%
\providecommand \doibase [0]{https://doi.org/}%
\providecommand \selectlanguage [0]{\@gobble}%
\providecommand \bibinfo  [0]{\@secondoftwo}%
\providecommand \bibfield  [0]{\@secondoftwo}%
\providecommand \translation [1]{[#1]}%
\providecommand \BibitemOpen [0]{}%
\providecommand \bibitemStop [0]{}%
\providecommand \bibitemNoStop [0]{.\EOS\space}%
\providecommand \EOS [0]{\spacefactor3000\relax}%
\providecommand \BibitemShut  [1]{\csname bibitem#1\endcsname}%
\let\auto@bib@innerbib\@empty
%</preamble>
\bibitem [{\citenamefont {Weinberg}(2008)}]{weinberg:2008}%
  \BibitemOpen
  \bibfield  {author} {\bibinfo {author} {\bibfnamefont {S.}~\bibnamefont
  {Weinberg}},\ }\href@noop {} {\emph {\bibinfo {title} {{Cosmology}}}}\
  (\bibinfo  {publisher} {Oxford University Press, New York},\ \bibinfo {year}
  {2008})\BibitemShut {NoStop}%
\bibitem [{\citenamefont {Abdalla}\ \emph {et~al.}(2022)\citenamefont {Abdalla}
  \emph {et~al.}}]{abdalla/etal:2022}%
  \BibitemOpen
  \bibfield  {author} {\bibinfo {author} {\bibfnamefont {E.}~\bibnamefont
  {Abdalla}} \emph {et~al.},\ }\href
  {https://doi.org/10.1016/j.jheap.2022.04.002} {\bibfield  {journal} {\bibinfo
   {journal} {J. High Energy Astrophys.}\ }\textbf {\bibinfo {volume} {34}},\
  \bibinfo {pages} {49} (\bibinfo {year} {2022})},\ \Eprint
  {https://arxiv.org/abs/2203.06142} {arXiv:2203.06142 [astro-ph.CO]}
  \BibitemShut {NoStop}%
\bibitem [{\citenamefont {Kamionkowski}\ and\ \citenamefont
  {Riess}(2022)}]{kamionkowski/riess:prep}%
  \BibitemOpen
  \bibfield  {author} {\bibinfo {author} {\bibfnamefont {M.}~\bibnamefont
  {Kamionkowski}}\ and\ \bibinfo {author} {\bibfnamefont {A.~G.}\ \bibnamefont
  {Riess}},\ }\href@noop {} {\bibfield  {journal} {\bibinfo  {journal} {arXiv
  preprint}\ } (\bibinfo {year} {2022})},\ \Eprint
  {https://arxiv.org/abs/2211.04492} {arXiv:2211.04492 [astro-ph.CO]}
  \BibitemShut {NoStop}%
\bibitem [{\citenamefont {Poulin}\ \emph {et~al.}(2023)\citenamefont {Poulin},
  \citenamefont {Smith},\ and\ \citenamefont {Karwal}}]{poulin/etal:prep}%
  \BibitemOpen
  \bibfield  {author} {\bibinfo {author} {\bibfnamefont {V.}~\bibnamefont
  {Poulin}}, \bibinfo {author} {\bibfnamefont {T.~L.}\ \bibnamefont {Smith}},\
  and\ \bibinfo {author} {\bibfnamefont {T.}~\bibnamefont {Karwal}},\
  }\href@noop {} {\bibfield  {journal} {\bibinfo  {journal} {arXiv preprint}\ }
  (\bibinfo {year} {2023})},\ \Eprint {https://arxiv.org/abs/2302.09032}
  {arXiv:2302.09032 [astro-ph.CO]} \BibitemShut {NoStop}%
\bibitem [{\citenamefont {Komatsu}(2022)}]{komatsu:2022}%
  \BibitemOpen
  \bibfield  {author} {\bibinfo {author} {\bibfnamefont {E.}~\bibnamefont
  {Komatsu}},\ }\href {https://doi.org/10.1038/s42254-022-00452-4} {\bibfield
  {journal} {\bibinfo  {journal} {Nat. Rev. Phys.}\ }\textbf {\bibinfo {volume}
  {4}},\ \bibinfo {pages} {452} (\bibinfo {year} {2022})},\ \Eprint
  {https://arxiv.org/abs/2202.13919} {arXiv:2202.13919 [astro-ph.CO]}
  \BibitemShut {NoStop}%
\bibitem [{\citenamefont {Doran}\ \emph {et~al.}(2001)\citenamefont {Doran},
  \citenamefont {Lilley}, \citenamefont {Schwindt},\ and\ \citenamefont
  {Wetterich}}]{doran/etal:2001}%
  \BibitemOpen
  \bibfield  {author} {\bibinfo {author} {\bibfnamefont {M.}~\bibnamefont
  {Doran}}, \bibinfo {author} {\bibfnamefont {M.~J.}\ \bibnamefont {Lilley}},
  \bibinfo {author} {\bibfnamefont {J.}~\bibnamefont {Schwindt}},\ and\
  \bibinfo {author} {\bibfnamefont {C.}~\bibnamefont {Wetterich}},\ }\href
  {https://doi.org/10.1086/322253} {\bibfield  {journal} {\bibinfo  {journal}
  {Astrophys. J.}\ }\textbf {\bibinfo {volume} {559}},\ \bibinfo {pages} {501}
  (\bibinfo {year} {2001})},\ \Eprint {https://arxiv.org/abs/astro-ph/0012139}
  {arXiv:astro-ph/0012139} \BibitemShut {NoStop}%
\bibitem [{\citenamefont {Wetterich}(2004)}]{wetterich:2004}%
  \BibitemOpen
  \bibfield  {author} {\bibinfo {author} {\bibfnamefont {C.}~\bibnamefont
  {Wetterich}},\ }\href {https://doi.org/10.1016/j.physletb.2004.05.008}
  {\bibfield  {journal} {\bibinfo  {journal} {Phys. Lett. B}\ }\textbf
  {\bibinfo {volume} {594}},\ \bibinfo {pages} {17} (\bibinfo {year} {2004})},\
  \Eprint {https://arxiv.org/abs/astro-ph/0403289} {arXiv:astro-ph/0403289}
  \BibitemShut {NoStop}%
\bibitem [{\citenamefont {Doran}\ and\ \citenamefont
  {Robbers}(2006)}]{doran/robbers:2006}%
  \BibitemOpen
  \bibfield  {author} {\bibinfo {author} {\bibfnamefont {M.}~\bibnamefont
  {Doran}}\ and\ \bibinfo {author} {\bibfnamefont {G.}~\bibnamefont
  {Robbers}},\ }\href {https://doi.org/10.1088/1475-7516/2006/06/026}
  {\bibfield  {journal} {\bibinfo  {journal} {J. Cosmol. Astropart. Phys.}\
  }\textbf {\bibinfo {volume} {06}}\bibfield  {number} {\bibinfo  {number} {
  (2006)},\ \bibinfo {pages} {026}},\ }\Eprint
  {https://arxiv.org/abs/astro-ph/0601544} {arXiv:astro-ph/0601544}
  \BibitemShut {NoStop}%
\bibitem [{\citenamefont {Karwal}\ and\ \citenamefont
  {Kamionkowski}(2016)}]{karwal/kamionkowski:2016}%
  \BibitemOpen
  \bibfield  {author} {\bibinfo {author} {\bibfnamefont {T.}~\bibnamefont
  {Karwal}}\ and\ \bibinfo {author} {\bibfnamefont {M.}~\bibnamefont
  {Kamionkowski}},\ }\href {https://doi.org/10.1103/PhysRevD.94.103523}
  {\bibfield  {journal} {\bibinfo  {journal} {Phys. Rev. D}\ }\textbf {\bibinfo
  {volume} {94}},\ \bibinfo {pages} {103523} (\bibinfo {year} {2016})},\
  \Eprint {https://arxiv.org/abs/1608.01309} {arXiv:1608.01309 [astro-ph.CO]}
  \BibitemShut {NoStop}%
\bibitem [{\citenamefont {Poulin}\ \emph {et~al.}(2019)\citenamefont {Poulin},
  \citenamefont {Smith}, \citenamefont {Karwal},\ and\ \citenamefont
  {Kamionkowski}}]{poulin/etal:2019}%
  \BibitemOpen
  \bibfield  {author} {\bibinfo {author} {\bibfnamefont {V.}~\bibnamefont
  {Poulin}}, \bibinfo {author} {\bibfnamefont {T.~L.}\ \bibnamefont {Smith}},
  \bibinfo {author} {\bibfnamefont {T.}~\bibnamefont {Karwal}},\ and\ \bibinfo
  {author} {\bibfnamefont {M.}~\bibnamefont {Kamionkowski}},\ }\href
  {https://doi.org/10.1103/PhysRevLett.122.221301} {\bibfield  {journal}
  {\bibinfo  {journal} {Phys. Rev. Lett.}\ }\textbf {\bibinfo {volume} {122}},\
  \bibinfo {pages} {221301} (\bibinfo {year} {2019})},\ \Eprint
  {https://arxiv.org/abs/1811.04083} {arXiv:1811.04083 [astro-ph.CO]}
  \BibitemShut {NoStop}%
\bibitem [{\citenamefont {Riess}\ \emph {et~al.}(2022)\citenamefont {Riess}
  \emph {et~al.}}]{riess/etal:2022}%
  \BibitemOpen
  \bibfield  {author} {\bibinfo {author} {\bibfnamefont {A.~G.}\ \bibnamefont
  {Riess}} \emph {et~al.},\ }\href {https://doi.org/10.3847/2041-8213/ac5c5b}
  {\bibfield  {journal} {\bibinfo  {journal} {Astrophys. J. Lett.}\ }\textbf
  {\bibinfo {volume} {934}},\ \bibinfo {pages} {L7} (\bibinfo {year} {2022})},\
  \Eprint {https://arxiv.org/abs/2112.04510} {arXiv:2112.04510 [astro-ph.CO]}
  \BibitemShut {NoStop}%
\bibitem [{\citenamefont {Ni}(1977)}]{ni:1977}%
  \BibitemOpen
  \bibfield  {author} {\bibinfo {author} {\bibfnamefont {W.-T.}\ \bibnamefont
  {Ni}},\ }\href {https://doi.org/10.1103/PhysRevLett.38.301} {\bibfield
  {journal} {\bibinfo  {journal} {Phys. Rev. Lett.}\ }\textbf {\bibinfo
  {volume} {38}},\ \bibinfo {pages} {301} (\bibinfo {year} {1977})}\BibitemShut
  {NoStop}%
\bibitem [{\citenamefont {Turner}\ and\ \citenamefont
  {Widrow}(1988)}]{turner/widrow:1988}%
  \BibitemOpen
  \bibfield  {author} {\bibinfo {author} {\bibfnamefont {M.~S.}\ \bibnamefont
  {Turner}}\ and\ \bibinfo {author} {\bibfnamefont {L.~M.}\ \bibnamefont
  {Widrow}},\ }\href {https://doi.org/10.1103/PhysRevD.37.2743} {\bibfield
  {journal} {\bibinfo  {journal} {Phys. Rev. D}\ }\textbf {\bibinfo {volume}
  {37}},\ \bibinfo {pages} {2743} (\bibinfo {year} {1988})}\BibitemShut
  {NoStop}%
\bibitem [{\citenamefont {Marsh}(2016)}]{marsh:2016}%
  \BibitemOpen
  \bibfield  {author} {\bibinfo {author} {\bibfnamefont {D.~J.~E.}\
  \bibnamefont {Marsh}},\ }\href
  {https://doi.org/10.1016/j.physrep.2016.06.005} {\bibfield  {journal}
  {\bibinfo  {journal} {Phys. Rep.}\ }\textbf {\bibinfo {volume} {643}},\
  \bibinfo {pages} {1} (\bibinfo {year} {2016})},\ \Eprint
  {https://arxiv.org/abs/1510.07633} {arXiv:1510.07633 [astro-ph.CO]}
  \BibitemShut {NoStop}%
\bibitem [{\citenamefont {Capparelli}\ \emph {et~al.}(2020)\citenamefont
  {Capparelli}, \citenamefont {Caldwell},\ and\ \citenamefont
  {Melchiorri}}]{capparelli/caldwell/melchiorri:2020}%
  \BibitemOpen
  \bibfield  {author} {\bibinfo {author} {\bibfnamefont {L.~M.}\ \bibnamefont
  {Capparelli}}, \bibinfo {author} {\bibfnamefont {R.~R.}\ \bibnamefont
  {Caldwell}},\ and\ \bibinfo {author} {\bibfnamefont {A.}~\bibnamefont
  {Melchiorri}},\ }\href {https://doi.org/10.1103/PhysRevD.101.123529}
  {\bibfield  {journal} {\bibinfo  {journal} {Phys. Rev. D}\ }\textbf {\bibinfo
  {volume} {101}},\ \bibinfo {pages} {123529} (\bibinfo {year} {2020})},\
  \Eprint {https://arxiv.org/abs/1909.04621} {arXiv:1909.04621 [astro-ph.CO]}
  \BibitemShut {NoStop}%
\bibitem [{\citenamefont {Fujita}\ \emph {et~al.}(2021)\citenamefont {Fujita},
  \citenamefont {Murai}, \citenamefont {Nakatsuka},\ and\ \citenamefont
  {Tsujikawa}}]{fujita/etal:2021b}%
  \BibitemOpen
  \bibfield  {author} {\bibinfo {author} {\bibfnamefont {T.}~\bibnamefont
  {Fujita}}, \bibinfo {author} {\bibfnamefont {K.}~\bibnamefont {Murai}},
  \bibinfo {author} {\bibfnamefont {H.}~\bibnamefont {Nakatsuka}},\ and\
  \bibinfo {author} {\bibfnamefont {S.}~\bibnamefont {Tsujikawa}},\ }\href
  {https://doi.org/10.1103/PhysRevD.103.043509} {\bibfield  {journal} {\bibinfo
   {journal} {Phys. Rev. D}\ }\textbf {\bibinfo {volume} {103}},\ \bibinfo
  {pages} {043509} (\bibinfo {year} {2021})},\ \Eprint
  {https://arxiv.org/abs/2011.11894} {arXiv:2011.11894 [astro-ph.CO]}
  \BibitemShut {NoStop}%
\bibitem [{\citenamefont {Nakagawa}\ \emph {et~al.}(2023)\citenamefont
  {Nakagawa}, \citenamefont {Takahashi},\ and\ \citenamefont
  {Yin}}]{Nakagawa:2022knn}%
  \BibitemOpen
  \bibfield  {author} {\bibinfo {author} {\bibfnamefont {S.}~\bibnamefont
  {Nakagawa}}, \bibinfo {author} {\bibfnamefont {F.}~\bibnamefont
  {Takahashi}},\ and\ \bibinfo {author} {\bibfnamefont {W.}~\bibnamefont
  {Yin}},\ }\href {https://doi.org/10.1103/PhysRevD.107.063016} {\bibfield
  {journal} {\bibinfo  {journal} {Phys. Rev. D}\ }\textbf {\bibinfo {volume}
  {107}},\ \bibinfo {pages} {063016} (\bibinfo {year} {2023})},\ \Eprint
  {https://arxiv.org/abs/2209.01107} {arXiv:2209.01107 [astro-ph.CO]}
  \BibitemShut {NoStop}%
\bibitem [{\citenamefont {Murai}\ \emph {et~al.}(2023)\citenamefont {Murai},
  \citenamefont {Naokawa}, \citenamefont {Namikawa},\ and\ \citenamefont
  {Komatsu}}]{Murai:2022zur}%
  \BibitemOpen
  \bibfield  {author} {\bibinfo {author} {\bibfnamefont {K.}~\bibnamefont
  {Murai}}, \bibinfo {author} {\bibfnamefont {F.}~\bibnamefont {Naokawa}},
  \bibinfo {author} {\bibfnamefont {T.}~\bibnamefont {Namikawa}},\ and\
  \bibinfo {author} {\bibfnamefont {E.}~\bibnamefont {Komatsu}},\ }\href
  {https://doi.org/10.1103/PhysRevD.107.L041302} {\bibfield  {journal}
  {\bibinfo  {journal} {Phys. Rev. D}\ }\textbf {\bibinfo {volume} {107}},\
  \bibinfo {pages} {L041302} (\bibinfo {year} {2023})},\ \Eprint
  {https://arxiv.org/abs/2209.07804} {arXiv:2209.07804 [astro-ph.CO]}
  \BibitemShut {NoStop}%
\bibitem [{\citenamefont {Greco}\ \emph {et~al.}(2023)\citenamefont {Greco},
  \citenamefont {Bartolo},\ and\ \citenamefont {Gruppuso}}]{Greco:2022xwj}%
  \BibitemOpen
  \bibfield  {author} {\bibinfo {author} {\bibfnamefont {A.}~\bibnamefont
  {Greco}}, \bibinfo {author} {\bibfnamefont {N.}~\bibnamefont {Bartolo}},\
  and\ \bibinfo {author} {\bibfnamefont {A.}~\bibnamefont {Gruppuso}},\ }\href
  {https://doi.org/10.1088/1475-7516/2023/05/026} {\bibfield  {journal}
  {\bibinfo  {journal} {J. Cosmol. Astropart. Phys.}\ }\textbf {\bibinfo
  {volume} {05}}\bibfield  {number} {\bibinfo  {number} { (2023)},\ \bibinfo
  {pages} {026}},\ }\Eprint {https://arxiv.org/abs/2211.06380}
  {arXiv:2211.06380 [astro-ph.CO]} \BibitemShut {NoStop}%
\bibitem [{\citenamefont {Galaverni}\ \emph {et~al.}(2023)\citenamefont
  {Galaverni}, \citenamefont {Finelli},\ and\ \citenamefont
  {Paoletti}}]{galaverni/finelli/paoletti:2023}%
  \BibitemOpen
  \bibfield  {author} {\bibinfo {author} {\bibfnamefont {M.}~\bibnamefont
  {Galaverni}}, \bibinfo {author} {\bibfnamefont {F.}~\bibnamefont {Finelli}},\
  and\ \bibinfo {author} {\bibfnamefont {D.}~\bibnamefont {Paoletti}},\ }\href
  {https://doi.org/10.1103/PhysRevD.107.083529} {\bibfield  {journal} {\bibinfo
   {journal} {Phys. Rev. D}\ }\textbf {\bibinfo {volume} {107}},\ \bibinfo
  {pages} {083529} (\bibinfo {year} {2023})},\ \Eprint
  {https://arxiv.org/abs/2301.07971} {arXiv:2301.07971 [astro-ph.CO]}
  \BibitemShut {NoStop}%
\bibitem [{\citenamefont {Smith}\ \emph {et~al.}(2020)\citenamefont {Smith},
  \citenamefont {Poulin},\ and\ \citenamefont {Amin}}]{smith/poulin/amin:2020}%
  \BibitemOpen
  \bibfield  {author} {\bibinfo {author} {\bibfnamefont {T.~L.}\ \bibnamefont
  {Smith}}, \bibinfo {author} {\bibfnamefont {V.}~\bibnamefont {Poulin}},\ and\
  \bibinfo {author} {\bibfnamefont {M.~A.}\ \bibnamefont {Amin}},\ }\href
  {https://doi.org/10.1103/PhysRevD.101.063523} {\bibfield  {journal} {\bibinfo
   {journal} {Phys. Rev. D}\ }\textbf {\bibinfo {volume} {101}},\ \bibinfo
  {pages} {063523} (\bibinfo {year} {2020})},\ \Eprint
  {https://arxiv.org/abs/1908.06995} {arXiv:1908.06995 [astro-ph.CO]}
  \BibitemShut {NoStop}%
\bibitem [{\citenamefont {Smith}\ \emph {et~al.}(2021)\citenamefont {Smith},
  \citenamefont {Poulin}, \citenamefont {Bernal}, \citenamefont {Boddy},
  \citenamefont {Kamionkowski},\ and\ \citenamefont {Murgia}}]{smith/eal:2021}%
  \BibitemOpen
  \bibfield  {author} {\bibinfo {author} {\bibfnamefont {T.~L.}\ \bibnamefont
  {Smith}}, \bibinfo {author} {\bibfnamefont {V.}~\bibnamefont {Poulin}},
  \bibinfo {author} {\bibfnamefont {J.~L.}\ \bibnamefont {Bernal}}, \bibinfo
  {author} {\bibfnamefont {K.~K.}\ \bibnamefont {Boddy}}, \bibinfo {author}
  {\bibfnamefont {M.}~\bibnamefont {Kamionkowski}},\ and\ \bibinfo {author}
  {\bibfnamefont {R.}~\bibnamefont {Murgia}},\ }\href
  {https://doi.org/10.1103/PhysRevD.103.123542} {\bibfield  {journal} {\bibinfo
   {journal} {Phys. Rev. D}\ }\textbf {\bibinfo {volume} {103}},\ \bibinfo
  {pages} {123542} (\bibinfo {year} {2021})},\ \Eprint
  {https://arxiv.org/abs/2009.10740} {arXiv:2009.10740 [astro-ph.CO]}
  \BibitemShut {NoStop}%
\bibitem [{\citenamefont {Murgia}\ \emph {et~al.}(2021)\citenamefont {Murgia},
  \citenamefont {Abell\'an},\ and\ \citenamefont
  {Poulin}}]{murgia/abellan/poulin:2021}%
  \BibitemOpen
  \bibfield  {author} {\bibinfo {author} {\bibfnamefont {R.}~\bibnamefont
  {Murgia}}, \bibinfo {author} {\bibfnamefont {G.~F.}\ \bibnamefont
  {Abell\'an}},\ and\ \bibinfo {author} {\bibfnamefont {V.}~\bibnamefont
  {Poulin}},\ }\href {https://doi.org/10.1103/PhysRevD.103.063502} {\bibfield
  {journal} {\bibinfo  {journal} {Phys. Rev. D}\ }\textbf {\bibinfo {volume}
  {103}},\ \bibinfo {pages} {063502} (\bibinfo {year} {2021})},\ \Eprint
  {https://arxiv.org/abs/2009.10733} {arXiv:2009.10733 [astro-ph.CO]}
  \BibitemShut {NoStop}%
\bibitem [{\citenamefont {Smith}\ \emph {et~al.}(2022)\citenamefont {Smith},
  \citenamefont {Lucca}, \citenamefont {Poulin}, \citenamefont {Abellan},
  \citenamefont {Balkenhol}, \citenamefont {Benabed}, \citenamefont {Galli},\
  and\ \citenamefont {Murgia}}]{smith/etal:2022}%
  \BibitemOpen
  \bibfield  {author} {\bibinfo {author} {\bibfnamefont {T.~L.}\ \bibnamefont
  {Smith}}, \bibinfo {author} {\bibfnamefont {M.}~\bibnamefont {Lucca}},
  \bibinfo {author} {\bibfnamefont {V.}~\bibnamefont {Poulin}}, \bibinfo
  {author} {\bibfnamefont {G.~F.}\ \bibnamefont {Abellan}}, \bibinfo {author}
  {\bibfnamefont {L.}~\bibnamefont {Balkenhol}}, \bibinfo {author}
  {\bibfnamefont {K.}~\bibnamefont {Benabed}}, \bibinfo {author} {\bibfnamefont
  {S.}~\bibnamefont {Galli}},\ and\ \bibinfo {author} {\bibfnamefont
  {R.}~\bibnamefont {Murgia}},\ }\href
  {https://doi.org/10.1103/PhysRevD.106.043526} {\bibfield  {journal} {\bibinfo
   {journal} {Phys. Rev. D}\ }\textbf {\bibinfo {volume} {106}},\ \bibinfo
  {pages} {043526} (\bibinfo {year} {2022})},\ \Eprint
  {https://arxiv.org/abs/2202.09379} {arXiv:2202.09379 [astro-ph.CO]}
  \BibitemShut {NoStop}%
\bibitem [{\citenamefont {Herold}\ \emph {et~al.}(2022)\citenamefont {Herold},
  \citenamefont {Ferreira},\ and\ \citenamefont
  {Komatsu}}]{herold/ferreira/komatsu:2022}%
  \BibitemOpen
  \bibfield  {author} {\bibinfo {author} {\bibfnamefont {L.}~\bibnamefont
  {Herold}}, \bibinfo {author} {\bibfnamefont {E.~G.~M.}\ \bibnamefont
  {Ferreira}},\ and\ \bibinfo {author} {\bibfnamefont {E.}~\bibnamefont
  {Komatsu}},\ }\href {https://doi.org/10.3847/2041-8213/ac63a3} {\bibfield
  {journal} {\bibinfo  {journal} {Astrophys. J. Lett.}\ }\textbf {\bibinfo
  {volume} {929}},\ \bibinfo {pages} {L16} (\bibinfo {year} {2022})},\ \Eprint
  {https://arxiv.org/abs/2112.12140} {arXiv:2112.12140 [astro-ph.CO]}
  \BibitemShut {NoStop}%
\bibitem [{\citenamefont {Simon}\ \emph {et~al.}(2023)\citenamefont {Simon},
  \citenamefont {Zhang}, \citenamefont {Poulin},\ and\ \citenamefont
  {Smith}}]{simon/etal:2023}%
  \BibitemOpen
  \bibfield  {author} {\bibinfo {author} {\bibfnamefont {T.}~\bibnamefont
  {Simon}}, \bibinfo {author} {\bibfnamefont {P.}~\bibnamefont {Zhang}},
  \bibinfo {author} {\bibfnamefont {V.}~\bibnamefont {Poulin}},\ and\ \bibinfo
  {author} {\bibfnamefont {T.~L.}\ \bibnamefont {Smith}},\ }\href
  {https://doi.org/10.1103/PhysRevD.107.063505} {\bibfield  {journal} {\bibinfo
   {journal} {Phys. Rev. D}\ }\textbf {\bibinfo {volume} {107}},\ \bibinfo
  {pages} {063505} (\bibinfo {year} {2023})},\ \Eprint
  {https://arxiv.org/abs/2208.05930} {arXiv:2208.05930 [astro-ph.CO]}
  \BibitemShut {NoStop}%
\bibitem [{\citenamefont {Herold}\ and\ \citenamefont
  {Ferreira}(2023)}]{herold/ferreira:2023}%
  \BibitemOpen
  \bibfield  {author} {\bibinfo {author} {\bibfnamefont {L.}~\bibnamefont
  {Herold}}\ and\ \bibinfo {author} {\bibfnamefont {E.~G.~M.}\ \bibnamefont
  {Ferreira}},\ }\href {https://doi.org/10.1103/PhysRevD.108.043513} {\bibfield
   {journal} {\bibinfo  {journal} {Phys. Rev. D}\ }\textbf {\bibinfo {volume}
  {108}},\ \bibinfo {pages} {043513} (\bibinfo {year} {2023})},\ \Eprint
  {https://arxiv.org/abs/2210.16296} {arXiv:2210.16296 [astro-ph.CO]}
  \BibitemShut {NoStop}%
\bibitem [{\citenamefont {Hill}\ \emph {et~al.}(2020)\citenamefont {Hill},
  \citenamefont {McDonough}, \citenamefont {Toomey},\ and\ \citenamefont
  {Alexander}}]{hill/etal:2020}%
  \BibitemOpen
  \bibfield  {author} {\bibinfo {author} {\bibfnamefont {J.~C.}\ \bibnamefont
  {Hill}}, \bibinfo {author} {\bibfnamefont {E.}~\bibnamefont {McDonough}},
  \bibinfo {author} {\bibfnamefont {M.~W.}\ \bibnamefont {Toomey}},\ and\
  \bibinfo {author} {\bibfnamefont {S.}~\bibnamefont {Alexander}},\ }\href
  {https://doi.org/10.1103/PhysRevD.102.043507} {\bibfield  {journal} {\bibinfo
   {journal} {Phys. Rev. D}\ }\textbf {\bibinfo {volume} {102}},\ \bibinfo
  {pages} {043507} (\bibinfo {year} {2020})},\ \Eprint
  {https://arxiv.org/abs/2003.07355} {arXiv:2003.07355 [astro-ph.CO]}
  \BibitemShut {NoStop}%
\bibitem [{\citenamefont {Ivanov}\ \emph {et~al.}(2020)\citenamefont {Ivanov},
  \citenamefont {McDonough}, \citenamefont {Hill}, \citenamefont {Simonovi\'c},
  \citenamefont {Toomey}, \citenamefont {Alexander},\ and\ \citenamefont
  {Zaldarriaga}}]{ivanov/etal:2020}%
  \BibitemOpen
  \bibfield  {author} {\bibinfo {author} {\bibfnamefont {M.~M.}\ \bibnamefont
  {Ivanov}}, \bibinfo {author} {\bibfnamefont {E.}~\bibnamefont {McDonough}},
  \bibinfo {author} {\bibfnamefont {J.~C.}\ \bibnamefont {Hill}}, \bibinfo
  {author} {\bibfnamefont {M.}~\bibnamefont {Simonovi\'c}}, \bibinfo {author}
  {\bibfnamefont {M.~W.}\ \bibnamefont {Toomey}}, \bibinfo {author}
  {\bibfnamefont {S.}~\bibnamefont {Alexander}},\ and\ \bibinfo {author}
  {\bibfnamefont {M.}~\bibnamefont {Zaldarriaga}},\ }\href
  {https://doi.org/10.1103/PhysRevD.102.103502} {\bibfield  {journal} {\bibinfo
   {journal} {Phys. Rev. D}\ }\textbf {\bibinfo {volume} {102}},\ \bibinfo
  {pages} {103502} (\bibinfo {year} {2020})},\ \Eprint
  {https://arxiv.org/abs/2006.11235} {arXiv:2006.11235 [astro-ph.CO]}
  \BibitemShut {NoStop}%
\bibitem [{\citenamefont {D'Amico}\ \emph {et~al.}(2021)\citenamefont
  {D'Amico}, \citenamefont {Senatore}, \citenamefont {Zhang},\ and\
  \citenamefont {Zheng}}]{damico/etal:2021}%
  \BibitemOpen
  \bibfield  {author} {\bibinfo {author} {\bibfnamefont {G.}~\bibnamefont
  {D'Amico}}, \bibinfo {author} {\bibfnamefont {L.}~\bibnamefont {Senatore}},
  \bibinfo {author} {\bibfnamefont {P.}~\bibnamefont {Zhang}},\ and\ \bibinfo
  {author} {\bibfnamefont {H.}~\bibnamefont {Zheng}},\ }\href
  {https://doi.org/10.1088/1475-7516/2021/05/072} {\bibfield  {journal}
  {\bibinfo  {journal} {J. Cosmol. Astropart. Phys.}\ }\textbf {\bibinfo
  {volume} {05}}\bibfield  {number} {\bibinfo  {number} { (2021)},\ \bibinfo
  {pages} {072}},\ }\Eprint {https://arxiv.org/abs/2006.12420}
  {arXiv:2006.12420 [astro-ph.CO]} \BibitemShut {NoStop}%
\bibitem [{\citenamefont {Hill}\ \emph {et~al.}(2022)\citenamefont {Hill} \emph
  {et~al.}}]{hill/etal:2022}%
  \BibitemOpen
  \bibfield  {author} {\bibinfo {author} {\bibfnamefont {J.~C.}\ \bibnamefont
  {Hill}} \emph {et~al.},\ }\href {https://doi.org/10.1103/PhysRevD.105.123536}
  {\bibfield  {journal} {\bibinfo  {journal} {Phys. Rev. D}\ }\textbf {\bibinfo
  {volume} {105}},\ \bibinfo {pages} {123536} (\bibinfo {year} {2022})},\
  \Eprint {https://arxiv.org/abs/2109.04451} {arXiv:2109.04451 [astro-ph.CO]}
  \BibitemShut {NoStop}%
\bibitem [{\citenamefont {La~Posta}\ \emph {et~al.}(2022)\citenamefont
  {La~Posta}, \citenamefont {Louis}, \citenamefont {Garrido},\ and\
  \citenamefont {Hill}}]{LaPosta:2021pgm}%
  \BibitemOpen
  \bibfield  {author} {\bibinfo {author} {\bibfnamefont {A.}~\bibnamefont
  {La~Posta}}, \bibinfo {author} {\bibfnamefont {T.}~\bibnamefont {Louis}},
  \bibinfo {author} {\bibfnamefont {X.}~\bibnamefont {Garrido}},\ and\ \bibinfo
  {author} {\bibfnamefont {J.~C.}\ \bibnamefont {Hill}},\ }\href
  {https://doi.org/10.1103/PhysRevD.105.083519} {\bibfield  {journal} {\bibinfo
   {journal} {Phys. Rev. D}\ }\textbf {\bibinfo {volume} {105}},\ \bibinfo
  {pages} {083519} (\bibinfo {year} {2022})},\ \Eprint
  {https://arxiv.org/abs/2112.10754} {arXiv:2112.10754 [astro-ph.CO]}
  \BibitemShut {NoStop}%
\bibitem [{\citenamefont {Reeves}\ \emph {et~al.}(2023)\citenamefont {Reeves},
  \citenamefont {Herold}, \citenamefont {Vagnozzi}, \citenamefont {Sherwin},\
  and\ \citenamefont {Ferreira}}]{reeves/etal:2023}%
  \BibitemOpen
  \bibfield  {author} {\bibinfo {author} {\bibfnamefont {A.}~\bibnamefont
  {Reeves}}, \bibinfo {author} {\bibfnamefont {L.}~\bibnamefont {Herold}},
  \bibinfo {author} {\bibfnamefont {S.}~\bibnamefont {Vagnozzi}}, \bibinfo
  {author} {\bibfnamefont {B.~D.}\ \bibnamefont {Sherwin}},\ and\ \bibinfo
  {author} {\bibfnamefont {E.~G.~M.}\ \bibnamefont {Ferreira}},\ }\href
  {https://doi.org/10.1093/mnras/stad317} {\bibfield  {journal} {\bibinfo
  {journal} {Mon. Not. Roy. Astron. Soc.}\ }\textbf {\bibinfo {volume} {520}},\
  \bibinfo {pages} {3688} (\bibinfo {year} {2023})},\ \Eprint
  {https://arxiv.org/abs/2207.01501} {arXiv:2207.01501 [astro-ph.CO]}
  \BibitemShut {NoStop}%
\bibitem [{\citenamefont {Cruz}\ \emph {et~al.}(2023)\citenamefont {Cruz},
  \citenamefont {Hannestad}, \citenamefont {Holm}, \citenamefont {Niedermann},
  \citenamefont {Sloth},\ and\ \citenamefont {Tram}}]{Cruz:2023cxy}%
  \BibitemOpen
  \bibfield  {author} {\bibinfo {author} {\bibfnamefont {J.~S.}\ \bibnamefont
  {Cruz}}, \bibinfo {author} {\bibfnamefont {S.}~\bibnamefont {Hannestad}},
  \bibinfo {author} {\bibfnamefont {E.~B.}\ \bibnamefont {Holm}}, \bibinfo
  {author} {\bibfnamefont {F.}~\bibnamefont {Niedermann}}, \bibinfo {author}
  {\bibfnamefont {M.~S.}\ \bibnamefont {Sloth}},\ and\ \bibinfo {author}
  {\bibfnamefont {T.}~\bibnamefont {Tram}},\ }\href@noop {} {\bibfield
  {journal} {\bibinfo  {journal} {arXiv preprint}\ } (\bibinfo {year}
  {2023})},\ \Eprint {https://arxiv.org/abs/2302.07934} {arXiv:2302.07934
  [astro-ph.CO]} \BibitemShut {NoStop}%
\bibitem [{\citenamefont {Goldstein}\ \emph {et~al.}(2023)\citenamefont
  {Goldstein}, \citenamefont {Hill}, \citenamefont {Ir\v{s}i\v{c}},\ and\
  \citenamefont {Sherwin}}]{goldstein/etal:prep}%
  \BibitemOpen
  \bibfield  {author} {\bibinfo {author} {\bibfnamefont {S.}~\bibnamefont
  {Goldstein}}, \bibinfo {author} {\bibfnamefont {J.~C.}\ \bibnamefont {Hill}},
  \bibinfo {author} {\bibfnamefont {V.}~\bibnamefont {Ir\v{s}i\v{c}}},\ and\
  \bibinfo {author} {\bibfnamefont {B.~D.}\ \bibnamefont {Sherwin}},\
  }\href@noop {} {\bibfield  {journal} {\bibinfo  {journal} {arXiv preprint}\ }
  (\bibinfo {year} {2023})},\ \Eprint {https://arxiv.org/abs/2303.00746}
  {arXiv:2303.00746 [astro-ph.CO]} \BibitemShut {NoStop}%
\bibitem [{\citenamefont {Zaldarriaga}\ and\ \citenamefont
  {Seljak}(1997)}]{zaldarriaga/seljak:1997}%
  \BibitemOpen
  \bibfield  {author} {\bibinfo {author} {\bibfnamefont {M.}~\bibnamefont
  {Zaldarriaga}}\ and\ \bibinfo {author} {\bibfnamefont {U.}~\bibnamefont
  {Seljak}},\ }\href {https://doi.org/10.1103/PhysRevD.55.1830} {\bibfield
  {journal} {\bibinfo  {journal} {Phys. Rev. D}\ }\textbf {\bibinfo {volume}
  {55}},\ \bibinfo {pages} {1830} (\bibinfo {year} {1997})},\ \Eprint
  {https://arxiv.org/abs/astro-ph/9609170} {arXiv:astro-ph/9609170}
  \BibitemShut {NoStop}%
\bibitem [{\citenamefont {Kamionkowski}\ \emph {et~al.}(1997)\citenamefont
  {Kamionkowski}, \citenamefont {Kosowsky},\ and\ \citenamefont
  {Stebbins}}]{kamionkowski/etal:1997}%
  \BibitemOpen
  \bibfield  {author} {\bibinfo {author} {\bibfnamefont {M.}~\bibnamefont
  {Kamionkowski}}, \bibinfo {author} {\bibfnamefont {A.}~\bibnamefont
  {Kosowsky}},\ and\ \bibinfo {author} {\bibfnamefont {A.}~\bibnamefont
  {Stebbins}},\ }\href {https://doi.org/10.1103/PhysRevD.55.7368} {\bibfield
  {journal} {\bibinfo  {journal} {Phys. Rev. D}\ }\textbf {\bibinfo {volume}
  {55}},\ \bibinfo {pages} {7368} (\bibinfo {year} {1997})},\ \Eprint
  {https://arxiv.org/abs/astro-ph/9611125} {arXiv:astro-ph/9611125}
  \BibitemShut {NoStop}%
\bibitem [{\citenamefont {Lue}\ \emph {et~al.}(1999)\citenamefont {Lue},
  \citenamefont {Wang},\ and\ \citenamefont
  {Kamionkowski}}]{lue/wang/kamionkowski:1999}%
  \BibitemOpen
  \bibfield  {author} {\bibinfo {author} {\bibfnamefont {A.}~\bibnamefont
  {Lue}}, \bibinfo {author} {\bibfnamefont {L.-M.}\ \bibnamefont {Wang}},\ and\
  \bibinfo {author} {\bibfnamefont {M.}~\bibnamefont {Kamionkowski}},\ }\href
  {https://doi.org/10.1103/PhysRevLett.83.1506} {\bibfield  {journal} {\bibinfo
   {journal} {Phys. Rev. Lett.}\ }\textbf {\bibinfo {volume} {83}},\ \bibinfo
  {pages} {1506} (\bibinfo {year} {1999})},\ \Eprint
  {https://arxiv.org/abs/astro-ph/9812088} {arXiv:astro-ph/9812088}
  \BibitemShut {NoStop}%
\bibitem [{\citenamefont {Eskilt}\ and\ \citenamefont
  {Komatsu}(2022)}]{Eskilt:2022cff}%
  \BibitemOpen
  \bibfield  {author} {\bibinfo {author} {\bibfnamefont {J.~R.}\ \bibnamefont
  {Eskilt}}\ and\ \bibinfo {author} {\bibfnamefont {E.}~\bibnamefont
  {Komatsu}},\ }\href {https://doi.org/10.1103/PhysRevD.106.063503} {\bibfield
  {journal} {\bibinfo  {journal} {Phys. Rev. D}\ }\textbf {\bibinfo {volume}
  {106}},\ \bibinfo {pages} {063503} (\bibinfo {year} {2022})},\ \Eprint
  {https://arxiv.org/abs/2205.13962} {arXiv:2205.13962 [astro-ph.CO]}
  \BibitemShut {NoStop}%
\bibitem [{\citenamefont {Carroll}\ \emph {et~al.}(1990)\citenamefont
  {Carroll}, \citenamefont {Field},\ and\ \citenamefont
  {Jackiw}}]{carroll/field/jackiw:1990}%
  \BibitemOpen
  \bibfield  {author} {\bibinfo {author} {\bibfnamefont {S.~M.}\ \bibnamefont
  {Carroll}}, \bibinfo {author} {\bibfnamefont {G.~B.}\ \bibnamefont {Field}},\
  and\ \bibinfo {author} {\bibfnamefont {R.}~\bibnamefont {Jackiw}},\ }\href
  {https://doi.org/10.1103/PhysRevD.41.1231} {\bibfield  {journal} {\bibinfo
  {journal} {Phys. Rev. D}\ }\textbf {\bibinfo {volume} {41}},\ \bibinfo
  {pages} {1231} (\bibinfo {year} {1990})}\BibitemShut {NoStop}%
\bibitem [{\citenamefont {Carroll}\ and\ \citenamefont
  {Field}(1991)}]{carroll/field:1991}%
  \BibitemOpen
  \bibfield  {author} {\bibinfo {author} {\bibfnamefont {S.~M.}\ \bibnamefont
  {Carroll}}\ and\ \bibinfo {author} {\bibfnamefont {G.~B.}\ \bibnamefont
  {Field}},\ }\href {https://doi.org/10.1103/PhysRevD.43.3789} {\bibfield
  {journal} {\bibinfo  {journal} {Phys. Rev. D}\ }\textbf {\bibinfo {volume}
  {43}},\ \bibinfo {pages} {3789} (\bibinfo {year} {1991})}\BibitemShut
  {NoStop}%
\bibitem [{\citenamefont {Harari}\ and\ \citenamefont
  {Sikivie}(1992)}]{harari/sikivie:1992}%
  \BibitemOpen
  \bibfield  {author} {\bibinfo {author} {\bibfnamefont {D.}~\bibnamefont
  {Harari}}\ and\ \bibinfo {author} {\bibfnamefont {P.}~\bibnamefont
  {Sikivie}},\ }\href {https://doi.org/10.1016/0370-2693(92)91363-E} {\bibfield
   {journal} {\bibinfo  {journal} {Phys. Lett. B}\ }\textbf {\bibinfo {volume}
  {289}},\ \bibinfo {pages} {67} (\bibinfo {year} {1992})}\BibitemShut
  {NoStop}%
\bibitem [{\citenamefont {Contreras}\ \emph {et~al.}(2017)\citenamefont
  {Contreras}, \citenamefont {Boubel},\ and\ \citenamefont
  {Scott}}]{contreras/boubel/scott:2017}%
  \BibitemOpen
  \bibfield  {author} {\bibinfo {author} {\bibfnamefont {D.}~\bibnamefont
  {Contreras}}, \bibinfo {author} {\bibfnamefont {P.}~\bibnamefont {Boubel}},\
  and\ \bibinfo {author} {\bibfnamefont {D.}~\bibnamefont {Scott}},\ }\href
  {https://doi.org/10.1088/1475-7516/2017/12/046} {\bibfield  {journal}
  {\bibinfo  {journal} {J. Cosmol. Astropart. Phys.}\ }\textbf {\bibinfo
  {volume} {12}}\bibfield  {number} {\bibinfo  {number} { (2017)},\ \bibinfo
  {pages} {046}},\ }\Eprint {https://arxiv.org/abs/1705.06387}
  {arXiv:1705.06387 [astro-ph.CO]} \BibitemShut {NoStop}%
\bibitem [{\citenamefont {Namikawa}\ \emph {et~al.}(2020)\citenamefont
  {Namikawa} \emph {et~al.}}]{namikawa/etal:2020}%
  \BibitemOpen
  \bibfield  {author} {\bibinfo {author} {\bibfnamefont {T.}~\bibnamefont
  {Namikawa}} \emph {et~al.} (\bibinfo {collaboration} {ACT Collaboration}),\
  }\href {https://doi.org/10.1103/PhysRevD.101.083527} {\bibfield  {journal}
  {\bibinfo  {journal} {Phys. Rev. D}\ }\textbf {\bibinfo {volume} {101}},\
  \bibinfo {pages} {083527} (\bibinfo {year} {2020})},\ \Eprint
  {https://arxiv.org/abs/2001.10465} {arXiv:2001.10465 [astro-ph.CO]}
  \BibitemShut {NoStop}%
\bibitem [{\citenamefont {Bianchini}\ \emph {et~al.}(2020)\citenamefont
  {Bianchini} \emph {et~al.}}]{bianchini/etal:2020}%
  \BibitemOpen
  \bibfield  {author} {\bibinfo {author} {\bibfnamefont {F.}~\bibnamefont
  {Bianchini}} \emph {et~al.} (\bibinfo {collaboration} {SPT Collaboration}),\
  }\href {https://doi.org/10.1103/PhysRevD.102.083504} {\bibfield  {journal}
  {\bibinfo  {journal} {Phys. Rev. D}\ }\textbf {\bibinfo {volume} {102}},\
  \bibinfo {pages} {083504} (\bibinfo {year} {2020})},\ \Eprint
  {https://arxiv.org/abs/2006.08061} {arXiv:2006.08061 [astro-ph.CO]}
  \BibitemShut {NoStop}%
\bibitem [{\citenamefont {Gruppuso}\ \emph {et~al.}(2020)\citenamefont
  {Gruppuso}, \citenamefont {Molinari}, \citenamefont {Natoli},\ and\
  \citenamefont {Pagano}}]{gruppuso/etal:2020}%
  \BibitemOpen
  \bibfield  {author} {\bibinfo {author} {\bibfnamefont {A.}~\bibnamefont
  {Gruppuso}}, \bibinfo {author} {\bibfnamefont {D.}~\bibnamefont {Molinari}},
  \bibinfo {author} {\bibfnamefont {P.}~\bibnamefont {Natoli}},\ and\ \bibinfo
  {author} {\bibfnamefont {L.}~\bibnamefont {Pagano}},\ }\href
  {https://doi.org/10.1088/1475-7516/2020/11/066} {\bibfield  {journal}
  {\bibinfo  {journal} {J. Cosmol. Astropart. Phys.}\ }\textbf {\bibinfo
  {volume} {11}}\bibfield  {number} {\bibinfo  {number} { (2020)},\ \bibinfo
  {pages} {066}},\ }\Eprint {https://arxiv.org/abs/2008.10334}
  {arXiv:2008.10334 [astro-ph.CO]} \BibitemShut {NoStop}%
\bibitem [{\citenamefont {Bortolami}\ \emph {et~al.}(2022)\citenamefont
  {Bortolami}, \citenamefont {Billi}, \citenamefont {Gruppuso}, \citenamefont
  {Natoli},\ and\ \citenamefont {Pagano}}]{bortolami/etal:2022}%
  \BibitemOpen
  \bibfield  {author} {\bibinfo {author} {\bibfnamefont {M.}~\bibnamefont
  {Bortolami}}, \bibinfo {author} {\bibfnamefont {M.}~\bibnamefont {Billi}},
  \bibinfo {author} {\bibfnamefont {A.}~\bibnamefont {Gruppuso}}, \bibinfo
  {author} {\bibfnamefont {P.}~\bibnamefont {Natoli}},\ and\ \bibinfo {author}
  {\bibfnamefont {L.}~\bibnamefont {Pagano}},\ }\href
  {https://doi.org/10.1088/1475-7516/2022/09/075} {\bibfield  {journal}
  {\bibinfo  {journal} {J. Cosmol. Astropart. Phys.}\ }\textbf {\bibinfo
  {volume} {09}}\bibfield  {number} {\bibinfo  {number} { (2022)},\ \bibinfo
  {pages} {075}},\ }\Eprint {https://arxiv.org/abs/2206.01635}
  {arXiv:2206.01635 [astro-ph.CO]} \BibitemShut {NoStop}%
\bibitem [{\citenamefont {Nakatsuka}\ \emph {et~al.}(2022)\citenamefont
  {Nakatsuka}, \citenamefont {Namikawa},\ and\ \citenamefont
  {Komatsu}}]{Nakatsuka:2022epj}%
  \BibitemOpen
  \bibfield  {author} {\bibinfo {author} {\bibfnamefont {H.}~\bibnamefont
  {Nakatsuka}}, \bibinfo {author} {\bibfnamefont {T.}~\bibnamefont
  {Namikawa}},\ and\ \bibinfo {author} {\bibfnamefont {E.}~\bibnamefont
  {Komatsu}},\ }\href {https://doi.org/10.1103/PhysRevD.105.123509} {\bibfield
  {journal} {\bibinfo  {journal} {Phys. Rev. D}\ }\textbf {\bibinfo {volume}
  {105}},\ \bibinfo {pages} {123509} (\bibinfo {year} {2022})},\ \Eprint
  {https://arxiv.org/abs/2203.08560} {arXiv:2203.08560 [astro-ph.CO]}
  \BibitemShut {NoStop}%
\bibitem [{\citenamefont {Feng}\ \emph {et~al.}(2005)\citenamefont {Feng},
  \citenamefont {Li}, \citenamefont {Li},\ and\ \citenamefont
  {Zhang}}]{feng/etal:2005}%
  \BibitemOpen
  \bibfield  {author} {\bibinfo {author} {\bibfnamefont {B.}~\bibnamefont
  {Feng}}, \bibinfo {author} {\bibfnamefont {H.}~\bibnamefont {Li}}, \bibinfo
  {author} {\bibfnamefont {M.}~\bibnamefont {Li}},\ and\ \bibinfo {author}
  {\bibfnamefont {X.}~\bibnamefont {Zhang}},\ }\href
  {https://doi.org/10.1016/j.physletb.2005.06.009} {\bibfield  {journal}
  {\bibinfo  {journal} {Phys. Lett. B}\ }\textbf {\bibinfo {volume} {620}},\
  \bibinfo {pages} {27} (\bibinfo {year} {2005})},\ \Eprint
  {https://arxiv.org/abs/hep-ph/0406269} {arXiv:hep-ph/0406269 [hep-ph]}
  \BibitemShut {NoStop}%
%%CITATION = HEP-PH/0406269;%%
\bibitem [{\citenamefont {Minami}\ and\ \citenamefont
  {Komatsu}(2020)}]{minami/komatsu:2020b}%
  \BibitemOpen
  \bibfield  {author} {\bibinfo {author} {\bibfnamefont {Y.}~\bibnamefont
  {Minami}}\ and\ \bibinfo {author} {\bibfnamefont {E.}~\bibnamefont
  {Komatsu}},\ }\href {https://doi.org/10.1103/PhysRevLett.125.221301}
  {\bibfield  {journal} {\bibinfo  {journal} {Phys. Rev. Lett.}\ }\textbf
  {\bibinfo {volume} {125}},\ \bibinfo {pages} {221301} (\bibinfo {year}
  {2020})},\ \Eprint {https://arxiv.org/abs/2011.11254} {arXiv:2011.11254
  [astro-ph.CO]} \BibitemShut {NoStop}%
\bibitem [{\citenamefont {Diego-Palazuelos}\ \emph {et~al.}(2022)\citenamefont
  {Diego-Palazuelos} \emph {et~al.}}]{NPIPE:2022}%
  \BibitemOpen
  \bibfield  {author} {\bibinfo {author} {\bibfnamefont {P.}~\bibnamefont
  {Diego-Palazuelos}} \emph {et~al.},\ }\href
  {https://doi.org/10.1103/PhysRevLett.128.091302} {\bibfield  {journal}
  {\bibinfo  {journal} {Phys. Rev. Lett.}\ }\textbf {\bibinfo {volume} {128}},\
  \bibinfo {pages} {091302} (\bibinfo {year} {2022})},\ \Eprint
  {https://arxiv.org/abs/2201.07682} {arXiv:2201.07682 [astro-ph.CO]}
  \BibitemShut {NoStop}%
\bibitem [{\citenamefont {Eskilt}(2022)}]{Eskilt:2022wav}%
  \BibitemOpen
  \bibfield  {author} {\bibinfo {author} {\bibfnamefont {J.~R.}\ \bibnamefont
  {Eskilt}},\ }\href {https://doi.org/10.1051/0004-6361/202243269} {\bibfield
  {journal} {\bibinfo  {journal} {Astron. Astrophys.}\ }\textbf {\bibinfo
  {volume} {662}},\ \bibinfo {pages} {A10} (\bibinfo {year} {2022})},\ \Eprint
  {https://arxiv.org/abs/2201.13347} {arXiv:2201.13347 [astro-ph.CO]}
  \BibitemShut {NoStop}%
\bibitem [{\citenamefont {Liu}\ \emph {et~al.}(2006)\citenamefont {Liu},
  \citenamefont {Lee},\ and\ \citenamefont {Ng}}]{liu/lee/ng:2006}%
  \BibitemOpen
  \bibfield  {author} {\bibinfo {author} {\bibfnamefont {G.-C.}\ \bibnamefont
  {Liu}}, \bibinfo {author} {\bibfnamefont {S.}~\bibnamefont {Lee}},\ and\
  \bibinfo {author} {\bibfnamefont {K.-W.}\ \bibnamefont {Ng}},\ }\href
  {https://doi.org/10.1103/PhysRevLett.97.161303} {\bibfield  {journal}
  {\bibinfo  {journal} {Phys. Rev. Lett.}\ }\textbf {\bibinfo {volume} {97}},\
  \bibinfo {pages} {161303} (\bibinfo {year} {2006})},\ \Eprint
  {https://arxiv.org/abs/astro-ph/0606248} {arXiv:astro-ph/0606248 [astro-ph]}
  \BibitemShut {NoStop}%
%%CITATION = ASTRO-PH/0606248;%%
\bibitem [{\citenamefont {Finelli}\ and\ \citenamefont
  {Galaverni}(2009)}]{finelli/galaverni:2009}%
  \BibitemOpen
  \bibfield  {author} {\bibinfo {author} {\bibfnamefont {F.}~\bibnamefont
  {Finelli}}\ and\ \bibinfo {author} {\bibfnamefont {M.}~\bibnamefont
  {Galaverni}},\ }\href {https://doi.org/10.1103/PhysRevD.79.063002} {\bibfield
   {journal} {\bibinfo  {journal} {Phys. Rev. D}\ }\textbf {\bibinfo {volume}
  {79}},\ \bibinfo {pages} {063002} (\bibinfo {year} {2009})},\ \Eprint
  {https://arxiv.org/abs/0802.4210} {arXiv:0802.4210 [astro-ph]} \BibitemShut
  {NoStop}%
\bibitem [{\citenamefont {Wu}\ \emph {et~al.}(2009)\citenamefont {Wu} \emph
  {et~al.}}]{wu/etal:2009}%
  \BibitemOpen
  \bibfield  {author} {\bibinfo {author} {\bibfnamefont {E.~Y.~S.}\
  \bibnamefont {Wu}} \emph {et~al.} (\bibinfo {collaboration} {QUaD
  Collaboration}),\ }\href {https://doi.org/10.1103/PhysRevLett.102.161302}
  {\bibfield  {journal} {\bibinfo  {journal} {Phys. Rev. Lett.}\ }\textbf
  {\bibinfo {volume} {102}},\ \bibinfo {pages} {161302} (\bibinfo {year}
  {2009})},\ \Eprint {https://arxiv.org/abs/0811.0618} {arXiv:0811.0618
  [astro-ph]} \BibitemShut {NoStop}%
%%CITATION = ARXIV:0811.0618;%%
\bibitem [{\citenamefont {Miller}\ \emph {et~al.}(2009)\citenamefont {Miller},
  \citenamefont {Shimon},\ and\ \citenamefont {Keating}}]{miller:2009}%
  \BibitemOpen
  \bibfield  {author} {\bibinfo {author} {\bibfnamefont {N.~J.}\ \bibnamefont
  {Miller}}, \bibinfo {author} {\bibfnamefont {M.}~\bibnamefont {Shimon}},\
  and\ \bibinfo {author} {\bibfnamefont {B.~G.}\ \bibnamefont {Keating}},\
  }\href {https://doi.org/10.1103/PhysRevD.79.103002} {\bibfield  {journal}
  {\bibinfo  {journal} {Phys. Rev. D}\ }\textbf {\bibinfo {volume} {79}},\
  \bibinfo {pages} {103002} (\bibinfo {year} {2009})},\ \Eprint
  {https://arxiv.org/abs/0903.1116} {arXiv:0903.1116 [astro-ph.CO]}
  \BibitemShut {NoStop}%
\bibitem [{\citenamefont {Komatsu}\ \emph {et~al.}(2011)\citenamefont {Komatsu}
  \emph {et~al.}}]{WMAP:2011}%
  \BibitemOpen
  \bibfield  {author} {\bibinfo {author} {\bibfnamefont {E.}~\bibnamefont
  {Komatsu}} \emph {et~al.} (\bibinfo {collaboration} {WMAP Collaboration}),\
  }\href {https://doi.org/10.1088/0067-0049/192/2/18} {\bibfield  {journal}
  {\bibinfo  {journal} {Astrophys. J. Suppl.}\ }\textbf {\bibinfo {volume}
  {192}},\ \bibinfo {pages} {18} (\bibinfo {year} {2011})},\ \Eprint
  {https://arxiv.org/abs/1001.4538} {arXiv:1001.4538 [astro-ph.CO]}
  \BibitemShut {NoStop}%
%%CITATION = ARXIV:1001.4538;%%
\bibitem [{\citenamefont {Krachmalnicoff}\ \emph {et~al.}(2022)\citenamefont
  {Krachmalnicoff} \emph {et~al.}}]{RAC:2022}%
  \BibitemOpen
  \bibfield  {author} {\bibinfo {author} {\bibfnamefont {N.}~\bibnamefont
  {Krachmalnicoff}} \emph {et~al.} (\bibinfo {collaboration} {LiteBIRD
  Collaboration}),\ }\href {https://doi.org/10.1088/1475-7516/2022/01/039}
  {\bibfield  {journal} {\bibinfo  {journal} {J. Cosmol. Astropart. Phys.}\
  }\textbf {\bibinfo {volume} {01}}\bibfield  {number} {\bibinfo  {number} {
  (2022)},\ \bibinfo {pages} {039}},\ }\Eprint
  {https://arxiv.org/abs/2111.09140} {arXiv:2111.09140 [astro-ph.CO]}
  \BibitemShut {NoStop}%
\bibitem [{\citenamefont {Cornelison}\ \emph {et~al.}(2022)\citenamefont
  {Cornelison} \emph {et~al.}}]{Cornelison:2022zrc}%
  \BibitemOpen
  \bibfield  {author} {\bibinfo {author} {\bibfnamefont {J.}~\bibnamefont
  {Cornelison}} \emph {et~al.},\ }\href {https://doi.org/10.1117/12.2620212}
  {\bibfield  {journal} {\bibinfo  {journal} {Proc. SPIE Int. Soc. Opt. Eng.}\
  }\textbf {\bibinfo {volume} {12190}},\ \bibinfo {pages} {121901X} (\bibinfo
  {year} {2022})},\ \Eprint {https://arxiv.org/abs/2207.14796}
  {arXiv:2207.14796 [astro-ph.IM]} \BibitemShut {NoStop}%
\bibitem [{\citenamefont {Minami}\ \emph {et~al.}(2019)\citenamefont {Minami},
  \citenamefont {Ochi}, \citenamefont {Ichiki}, \citenamefont {Katayama},
  \citenamefont {Komatsu},\ and\ \citenamefont {Matsumura}}]{minami/etal:2019}%
  \BibitemOpen
  \bibfield  {author} {\bibinfo {author} {\bibfnamefont {Y.}~\bibnamefont
  {Minami}}, \bibinfo {author} {\bibfnamefont {H.}~\bibnamefont {Ochi}},
  \bibinfo {author} {\bibfnamefont {K.}~\bibnamefont {Ichiki}}, \bibinfo
  {author} {\bibfnamefont {N.}~\bibnamefont {Katayama}}, \bibinfo {author}
  {\bibfnamefont {E.}~\bibnamefont {Komatsu}},\ and\ \bibinfo {author}
  {\bibfnamefont {T.}~\bibnamefont {Matsumura}},\ }\href
  {https://doi.org/10.1093/ptep/ptz079} {\bibfield  {journal} {\bibinfo
  {journal} {Prog. Theor. Exp. Phys.}\ }\textbf {\bibinfo {volume} {2019}},\
  \bibinfo {pages} {083E02} (\bibinfo {year} {2019})},\ \Eprint
  {https://arxiv.org/abs/1904.12440} {arXiv:1904.12440 [astro-ph.CO]}
  \BibitemShut {NoStop}%
\bibitem [{\citenamefont {{Sherwin}}\ and\ \citenamefont
  {{Namikawa}}(2023)}]{sherwin/namikawa:2023}%
  \BibitemOpen
  \bibfield  {author} {\bibinfo {author} {\bibfnamefont {B.~D.}\ \bibnamefont
  {{Sherwin}}}\ and\ \bibinfo {author} {\bibfnamefont {T.}~\bibnamefont
  {{Namikawa}}},\ }\href {https://doi.org/10.1093/mnras/stac3146} {\bibfield
  {journal} {\bibinfo  {journal} {Mon. Not. Roy. Astron. Soc.}\ }\textbf
  {\bibinfo {volume} {520}},\ \bibinfo {pages} {3298} (\bibinfo {year}
  {2023})},\ \Eprint {https://arxiv.org/abs/2108.09287} {arXiv:2108.09287
  [astro-ph.CO]} \BibitemShut {NoStop}%
\bibitem [{\citenamefont {Diego-Palazuelos}\ \emph {et~al.}(2023)\citenamefont
  {Diego-Palazuelos} \emph {et~al.}}]{diego-palazuelos/etal:2023}%
  \BibitemOpen
  \bibfield  {author} {\bibinfo {author} {\bibfnamefont {P.}~\bibnamefont
  {Diego-Palazuelos}} \emph {et~al.},\ }\href
  {https://doi.org/10.1088/1475-7516/2023/01/044} {\bibfield  {journal}
  {\bibinfo  {journal} {J. Cosmol. Astropart. Phys.}\ }\textbf {\bibinfo
  {volume} {01}}\bibfield  {number} {\bibinfo  {number} { (2023)},\ \bibinfo
  {pages} {044}},\ }\Eprint {https://arxiv.org/abs/2210.07655}
  {arXiv:2210.07655 [astro-ph.CO]} \BibitemShut {NoStop}%
\bibitem [{\citenamefont {{Planck Collaboration III}}(2020)}]{Planck2018III}%
  \BibitemOpen
  \bibfield  {author} {\bibinfo {author} {\bibnamefont {{Planck Collaboration
  III}}},\ }\href {https://doi.org/10.1051/0004-6361/201832909} {\bibfield
  {journal} {\bibinfo  {journal} {Astron. Astrophys.}\ }\textbf {\bibinfo
  {volume} {641}},\ \bibinfo {pages} {A3} (\bibinfo {year} {2020})},\ \Eprint
  {https://arxiv.org/abs/1807.06207} {arXiv:1807.06207} \BibitemShut {NoStop}%
\bibitem [{\citenamefont {{Planck Collaboration Int.
  LVII}}(2020)}]{PlanckIntLVII}%
  \BibitemOpen
  \bibfield  {author} {\bibinfo {author} {\bibnamefont {{Planck Collaboration
  Int. LVII}}},\ }\href {https://doi.org/10.1051/0004-6361/202038073}
  {\bibfield  {journal} {\bibinfo  {journal} {Astron. Astrophys.}\ }\textbf
  {\bibinfo {volume} {643}},\ \bibinfo {pages} {A42} (\bibinfo {year}
  {2020})},\ \Eprint {https://arxiv.org/abs/2007.04997} {arXiv:2007.04997}
  \BibitemShut {NoStop}%
\bibitem [{\citenamefont {Chon}\ \emph {et~al.}(2004)\citenamefont {Chon},
  \citenamefont {Challinor}, \citenamefont {Prunet}, \citenamefont {Hivon},\
  and\ \citenamefont {Szapudi}}]{Chon:2003gx}%
  \BibitemOpen
  \bibfield  {author} {\bibinfo {author} {\bibfnamefont {G.}~\bibnamefont
  {Chon}}, \bibinfo {author} {\bibfnamefont {A.}~\bibnamefont {Challinor}},
  \bibinfo {author} {\bibfnamefont {S.}~\bibnamefont {Prunet}}, \bibinfo
  {author} {\bibfnamefont {E.}~\bibnamefont {Hivon}},\ and\ \bibinfo {author}
  {\bibfnamefont {I.}~\bibnamefont {Szapudi}},\ }\href
  {https://doi.org/10.1111/j.1365-2966.2004.07737.x} {\bibfield  {journal}
  {\bibinfo  {journal} {Mon. Not. Roy. Astron. Soc.}\ }\textbf {\bibinfo
  {volume} {350}},\ \bibinfo {pages} {914} (\bibinfo {year} {2004})},\ \Eprint
  {https://arxiv.org/abs/astro-ph/0303414} {arXiv:astro-ph/0303414}
  \BibitemShut {NoStop}%
\bibitem [{\citenamefont {G\'orski}\ \emph {et~al.}(2005)\citenamefont
  {G\'orski}, \citenamefont {Hivon}, \citenamefont {Banday}, \citenamefont
  {Wandelt}, \citenamefont {Hansen}, \citenamefont {Reinecke},\ and\
  \citenamefont {Bartelman}}]{gorski/etal:2005}%
  \BibitemOpen
  \bibfield  {author} {\bibinfo {author} {\bibfnamefont {K.~M.}\ \bibnamefont
  {G\'orski}}, \bibinfo {author} {\bibfnamefont {E.}~\bibnamefont {Hivon}},
  \bibinfo {author} {\bibfnamefont {A.~J.}\ \bibnamefont {Banday}}, \bibinfo
  {author} {\bibfnamefont {B.~D.}\ \bibnamefont {Wandelt}}, \bibinfo {author}
  {\bibfnamefont {F.~K.}\ \bibnamefont {Hansen}}, \bibinfo {author}
  {\bibfnamefont {M.}~\bibnamefont {Reinecke}},\ and\ \bibinfo {author}
  {\bibfnamefont {M.}~\bibnamefont {Bartelman}},\ }\href
  {https://doi.org/10.1086/427976} {\bibfield  {journal} {\bibinfo  {journal}
  {Astrophys. J.}\ }\textbf {\bibinfo {volume} {622}},\ \bibinfo {pages} {759}
  (\bibinfo {year} {2005})},\ \Eprint {https://arxiv.org/abs/astro-ph/0409513}
  {arXiv:astro-ph/0409513} \BibitemShut {NoStop}%
\bibitem [{\citenamefont {{Planck Collaboration Int.
  XLIX}}(2016)}]{PlanckIntXLIX}%
  \BibitemOpen
  \bibfield  {author} {\bibinfo {author} {\bibnamefont {{Planck Collaboration
  Int. XLIX}}},\ }\href {https://doi.org/10.1051/0004-6361/201629018}
  {\bibfield  {journal} {\bibinfo  {journal} {Astron. Astrophys.}\ }\textbf
  {\bibinfo {volume} {596}},\ \bibinfo {pages} {A110} (\bibinfo {year}
  {2016})},\ \Eprint {https://arxiv.org/abs/1605.08633} {arXiv:1605.08633}
  \BibitemShut {NoStop}%
\bibitem [{\citenamefont {Foreman-Mackey}\ \emph {et~al.}(2013)\citenamefont
  {Foreman-Mackey}, \citenamefont {Hogg}, \citenamefont {Lang},\ and\
  \citenamefont {Goodman}}]{ForemanMackey:2012ig}%
  \BibitemOpen
  \bibfield  {author} {\bibinfo {author} {\bibfnamefont {D.}~\bibnamefont
  {Foreman-Mackey}}, \bibinfo {author} {\bibfnamefont {D.~W.}\ \bibnamefont
  {Hogg}}, \bibinfo {author} {\bibfnamefont {D.}~\bibnamefont {Lang}},\ and\
  \bibinfo {author} {\bibfnamefont {J.}~\bibnamefont {Goodman}},\ }\href
  {https://doi.org/10.1086/670067} {\bibfield  {journal} {\bibinfo  {journal}
  {Publ. Astron. Soc. Pac.}\ }\textbf {\bibinfo {volume} {125}},\ \bibinfo
  {pages} {306} (\bibinfo {year} {2013})},\ \Eprint
  {https://arxiv.org/abs/1202.3665} {arXiv:1202.3665 [astro-ph.IM]}
  \BibitemShut {NoStop}%
\bibitem [{\citenamefont {{Lewis}}\ \emph {et~al.}(2000)\citenamefont
  {{Lewis}}, \citenamefont {{Challinor}},\ and\ \citenamefont
  {{Lasenby}}}]{Lewis:2000}%
  \BibitemOpen
  \bibfield  {author} {\bibinfo {author} {\bibfnamefont {A.}~\bibnamefont
  {{Lewis}}}, \bibinfo {author} {\bibfnamefont {A.}~\bibnamefont
  {{Challinor}}},\ and\ \bibinfo {author} {\bibfnamefont {A.}~\bibnamefont
  {{Lasenby}}},\ }\href {https://doi.org/10.1086/309179} {\bibfield  {journal}
  {\bibinfo  {journal} {Astrophys. J.}\ }\textbf {\bibinfo {volume} {538}},\
  \bibinfo {pages} {473} (\bibinfo {year} {2000})},\ \Eprint
  {https://arxiv.org/abs/astro-ph/9911177} {astro-ph/9911177} \BibitemShut
  {NoStop}%
\bibitem [{\citenamefont {{Planck Collaboration VI}}(2020)}]{Planck2018VI}%
  \BibitemOpen
  \bibfield  {author} {\bibinfo {author} {\bibnamefont {{Planck Collaboration
  VI}}},\ }\href {https://doi.org/10.1051/0004-6361/201833910} {\bibfield
  {journal} {\bibinfo  {journal} {Astron. Astrophys.}\ }\textbf {\bibinfo
  {volume} {641}},\ \bibinfo {pages} {A6} (\bibinfo {year} {2020})},\ \Eprint
  {https://arxiv.org/abs/1807.06209} {arXiv:1807.06209} \BibitemShut {NoStop}%
\bibitem [{\citenamefont {Blas}\ \emph {et~al.}(2011)\citenamefont {Blas},
  \citenamefont {Lesgourgues},\ and\ \citenamefont {Tram}}]{Blas:2011rf}%
  \BibitemOpen
  \bibfield  {author} {\bibinfo {author} {\bibfnamefont {D.}~\bibnamefont
  {Blas}}, \bibinfo {author} {\bibfnamefont {J.}~\bibnamefont {Lesgourgues}},\
  and\ \bibinfo {author} {\bibfnamefont {T.}~\bibnamefont {Tram}},\ }\href
  {https://doi.org/10.1088/1475-7516/2011/07/034} {\bibfield  {journal}
  {\bibinfo  {journal} {J. Cosmol. Astropart. Phys.}\ }\textbf {\bibinfo
  {volume} {07}}\bibfield  {number} {\bibinfo  {number} { (2011)},\ \bibinfo
  {pages} {034}},\ }\Eprint {https://arxiv.org/abs/1104.2933} {arXiv:1104.2933
  [astro-ph.CO]} \BibitemShut {NoStop}%
\bibitem [{\citenamefont {Naokawa}\ and\ \citenamefont
  {Namikawa}(2023)}]{Naokawa:2023upt}%
  \BibitemOpen
  \bibfield  {author} {\bibinfo {author} {\bibfnamefont {F.}~\bibnamefont
  {Naokawa}}\ and\ \bibinfo {author} {\bibfnamefont {T.}~\bibnamefont
  {Namikawa}},\ }\href@noop {} {\bibfield  {journal} {\bibinfo  {journal}
  {arXiv preprint}\ } (\bibinfo {year} {2023})},\ \Eprint
  {https://arxiv.org/abs/2305.13976} {arXiv:2305.13976 [astro-ph.CO]}
  \BibitemShut {NoStop}%
\bibitem [{\citenamefont {{Planck Collaboration V}}(2020)}]{Planck2018V}%
  \BibitemOpen
  \bibfield  {author} {\bibinfo {author} {\bibnamefont {{Planck Collaboration
  V}}},\ }\href {https://doi.org/10.1051/0004-6361/201936386} {\bibfield
  {journal} {\bibinfo  {journal} {Astron. Astrophys.}\ }\textbf {\bibinfo
  {volume} {641}},\ \bibinfo {pages} {A5} (\bibinfo {year} {2020})},\ \Eprint
  {https://arxiv.org/abs/1907.12875} {arXiv:1907.12875} \BibitemShut {NoStop}%
\bibitem [{\citenamefont {Alam}\ \emph {et~al.}(2017)\citenamefont {Alam} \emph
  {et~al.}}]{BOSS:2016wmc}%
  \BibitemOpen
  \bibfield  {author} {\bibinfo {author} {\bibfnamefont {S.}~\bibnamefont
  {Alam}} \emph {et~al.} (\bibinfo {collaboration} {BOSS Collaboration}),\
  }\href {https://doi.org/10.1093/mnras/stx721} {\bibfield  {journal} {\bibinfo
   {journal} {Mon. Not. Roy. Astron. Soc.}\ }\textbf {\bibinfo {volume}
  {470}},\ \bibinfo {pages} {2617} (\bibinfo {year} {2017})},\ \Eprint
  {https://arxiv.org/abs/1607.03155} {arXiv:1607.03155 [astro-ph.CO]}
  \BibitemShut {NoStop}%
\bibitem [{\citenamefont {Rudelius}(2023)}]{rudelius:2023}%
  \BibitemOpen
  \bibfield  {author} {\bibinfo {author} {\bibfnamefont {T.}~\bibnamefont
  {Rudelius}},\ }\href {https://doi.org/10.1088/1475-7516/2023/01/014}
  {\bibfield  {journal} {\bibinfo  {journal} {J. Cosmol. Astropart. Phys.}\
  }\textbf {\bibinfo {volume} {01}}\bibfield  {number} {\bibinfo  {number} {
  (2023)},\ \bibinfo {pages} {014}},\ }\Eprint
  {https://arxiv.org/abs/2203.05575} {arXiv:2203.05575 [hep-th]} \BibitemShut
  {NoStop}%
\bibitem [{\citenamefont {McDonough}\ and\ \citenamefont
  {Scalisi}(2022)}]{mcdonough/scalisi:prep}%
  \BibitemOpen
  \bibfield  {author} {\bibinfo {author} {\bibfnamefont {E.}~\bibnamefont
  {McDonough}}\ and\ \bibinfo {author} {\bibfnamefont {M.}~\bibnamefont
  {Scalisi}},\ }\href@noop {} {\bibfield  {journal} {\bibinfo  {journal} {arXiv
  preprint}\ } (\bibinfo {year} {2022})},\ \Eprint
  {https://arxiv.org/abs/2209.00011} {arXiv:2209.00011 [hep-th]} \BibitemShut
  {NoStop}%
\bibitem [{\citenamefont {Cicoli}\ \emph {et~al.}(2023)\citenamefont {Cicoli},
  \citenamefont {Licheri}, \citenamefont {Mahanta}, \citenamefont {McDonough},
  \citenamefont {Pedro},\ and\ \citenamefont {Scalisi}}]{cicoli/etal:2023}%
  \BibitemOpen
  \bibfield  {author} {\bibinfo {author} {\bibfnamefont {M.}~\bibnamefont
  {Cicoli}}, \bibinfo {author} {\bibfnamefont {M.}~\bibnamefont {Licheri}},
  \bibinfo {author} {\bibfnamefont {R.}~\bibnamefont {Mahanta}}, \bibinfo
  {author} {\bibfnamefont {E.}~\bibnamefont {McDonough}}, \bibinfo {author}
  {\bibfnamefont {F.~G.}\ \bibnamefont {Pedro}},\ and\ \bibinfo {author}
  {\bibfnamefont {M.}~\bibnamefont {Scalisi}},\ }\href
  {https://doi.org/10.1007/JHEP06(2023)052} {\bibfield  {journal} {\bibinfo
  {journal} {J. High Energy Phys.}\ }\textbf {\bibinfo {volume} {06}}\bibfield
  {number} {\bibinfo  {number} { (2023)},\ \bibinfo {pages} {052}},\ }\Eprint
  {https://arxiv.org/abs/2303.03414} {arXiv:2303.03414 [hep-th]} \BibitemShut
  {NoStop}%
\bibitem [{\citenamefont {Arkani-Hamed}\ \emph {et~al.}(2007)\citenamefont
  {Arkani-Hamed}, \citenamefont {Motl}, \citenamefont {Nicolis},\ and\
  \citenamefont {Vafa}}]{Arkani-Hamed:2006emk}%
  \BibitemOpen
  \bibfield  {author} {\bibinfo {author} {\bibfnamefont {N.}~\bibnamefont
  {Arkani-Hamed}}, \bibinfo {author} {\bibfnamefont {L.}~\bibnamefont {Motl}},
  \bibinfo {author} {\bibfnamefont {A.}~\bibnamefont {Nicolis}},\ and\ \bibinfo
  {author} {\bibfnamefont {C.}~\bibnamefont {Vafa}},\ }\href
  {https://doi.org/10.1088/1126-6708/2007/06/060} {\bibfield  {journal}
  {\bibinfo  {journal} {J. High Energy Phys.}\ }\textbf {\bibinfo {volume}
  {06}}\bibfield  {number} {\bibinfo  {number} { (2007)},\ \bibinfo {pages}
  {060}},\ }\Eprint {https://arxiv.org/abs/hep-th/0601001}
  {arXiv:hep-th/0601001} \BibitemShut {NoStop}%
\bibitem [{\citenamefont {Harlow}\ \emph {et~al.}(2022)\citenamefont {Harlow},
  \citenamefont {Heidenreich}, \citenamefont {Reece},\ and\ \citenamefont
  {Rudelius}}]{Harlow:2022gzl}%
  \BibitemOpen
  \bibfield  {author} {\bibinfo {author} {\bibfnamefont {D.}~\bibnamefont
  {Harlow}}, \bibinfo {author} {\bibfnamefont {B.}~\bibnamefont {Heidenreich}},
  \bibinfo {author} {\bibfnamefont {M.}~\bibnamefont {Reece}},\ and\ \bibinfo
  {author} {\bibfnamefont {T.}~\bibnamefont {Rudelius}},\ }\href@noop {}
  {\bibfield  {journal} {\bibinfo  {journal} {arXiv preprint}\ } (\bibinfo
  {year} {2022})},\ \Eprint {https://arxiv.org/abs/2201.08380}
  {arXiv:2201.08380 [hep-th]} \BibitemShut {NoStop}%
\bibitem [{\citenamefont {Heidenreich}\ \emph {et~al.}(2021)\citenamefont
  {Heidenreich}, \citenamefont {Reece},\ and\ \citenamefont
  {Rudelius}}]{Heidenreich:2021yda}%
  \BibitemOpen
  \bibfield  {author} {\bibinfo {author} {\bibfnamefont {B.}~\bibnamefont
  {Heidenreich}}, \bibinfo {author} {\bibfnamefont {M.}~\bibnamefont {Reece}},\
  and\ \bibinfo {author} {\bibfnamefont {T.}~\bibnamefont {Rudelius}},\ }\href
  {https://doi.org/10.1007/JHEP11(2021)004} {\bibfield  {journal} {\bibinfo
  {journal} {J. High Energy Phys.}\ }\textbf {\bibinfo {volume} {11}}\bibfield
  {number} {\bibinfo  {number} { (2021)},\ \bibinfo {pages} {004}},\ }\Eprint
  {https://arxiv.org/abs/2108.11383} {arXiv:2108.11383 [hep-th]} \BibitemShut
  {NoStop}%
\bibitem [{\citenamefont {Kaya}\ and\ \citenamefont
  {Rudelius}(2022)}]{Kaya:2022edp}%
  \BibitemOpen
  \bibfield  {author} {\bibinfo {author} {\bibfnamefont {S.}~\bibnamefont
  {Kaya}}\ and\ \bibinfo {author} {\bibfnamefont {T.}~\bibnamefont
  {Rudelius}},\ }\href {https://doi.org/10.1007/JHEP07(2022)040} {\bibfield
  {journal} {\bibinfo  {journal} {J. High Energy Phys.}\ }\textbf {\bibinfo
  {volume} {07}}\bibfield  {number} {\bibinfo  {number} { (2022)},\ \bibinfo
  {pages} {040}},\ }\Eprint {https://arxiv.org/abs/2202.04655}
  {arXiv:2202.04655 [hep-th]} \BibitemShut {NoStop}%
\bibitem [{\citenamefont {Ade}\ \emph {et~al.}(2019)\citenamefont {Ade} \emph
  {et~al.}}]{SimonsObservatory:2019}%
  \BibitemOpen
  \bibfield  {author} {\bibinfo {author} {\bibfnamefont {P.}~\bibnamefont
  {Ade}} \emph {et~al.} (\bibinfo {collaboration} {Simons Observatory
  Collaboration}),\ }\href {https://doi.org/10.1088/1475-7516/2019/02/056}
  {\bibfield  {journal} {\bibinfo  {journal} {J. Cosmol. Astropart. Phys.}\
  }\textbf {\bibinfo {volume} {02}}\bibfield  {number} {\bibinfo  {number} {
  (2019)},\ \bibinfo {pages} {056}},\ }\Eprint
  {https://arxiv.org/abs/1808.07445} {arXiv:1808.07445 [astro-ph.CO]}
  \BibitemShut {NoStop}%
\bibitem [{\citenamefont {Abazajian}\ \emph {et~al.}(2019)\citenamefont
  {Abazajian} \emph {et~al.}}]{CMB-S4:2019}%
  \BibitemOpen
  \bibfield  {author} {\bibinfo {author} {\bibfnamefont {K.}~\bibnamefont
  {Abazajian}} \emph {et~al.} (\bibinfo {collaboration} {CMB-S4
  Collaboration}),\ }\href@noop {} {\bibfield  {journal} {\bibinfo  {journal}
  {arXiv preprint}\ } (\bibinfo {year} {2019})},\ \Eprint
  {https://arxiv.org/abs/1907.04473} {arXiv:1907.04473 [astro-ph.IM]}
  \BibitemShut {NoStop}%
\bibitem [{\citenamefont {Moncelsi}\ \emph {et~al.}(2020)\citenamefont
  {Moncelsi} \emph {et~al.}}]{SPO:2020}%
  \BibitemOpen
  \bibfield  {author} {\bibinfo {author} {\bibfnamefont {L.}~\bibnamefont
  {Moncelsi}} \emph {et~al.},\ }\href {https://doi.org/10.1117/12.2561995}
  {\bibfield  {journal} {\bibinfo  {journal} {Proc. SPIE Int. Soc. Opt. Eng.}\
  }\textbf {\bibinfo {volume} {11453}},\ \bibinfo {pages} {1145314} (\bibinfo
  {year} {2020})},\ \Eprint {https://arxiv.org/abs/2012.04047}
  {arXiv:2012.04047 [astro-ph.IM]} \BibitemShut {NoStop}%
\bibitem [{\citenamefont {Hanany}\ \emph {et~al.}(2019)\citenamefont {Hanany}
  \emph {et~al.}}]{NASAPICO:2019}%
  \BibitemOpen
  \bibfield  {author} {\bibinfo {author} {\bibfnamefont {S.}~\bibnamefont
  {Hanany}} \emph {et~al.} (\bibinfo {collaboration} {PICO Collaboration}),\
  }\href@noop {} {\bibfield  {journal} {\bibinfo  {journal} {arXiv preprint}\ }
  (\bibinfo {year} {2019})},\ \Eprint {https://arxiv.org/abs/1902.10541}
  {arXiv:1902.10541 [astro-ph.IM]} \BibitemShut {NoStop}%
\bibitem [{\citenamefont {{LiteBIRD Collaboration}}(2022)}]{LiteBIRD:2022}%
  \BibitemOpen
  \bibfield  {author} {\bibinfo {author} {\bibnamefont {{LiteBIRD
  Collaboration}}},\ }\href@noop {} {\bibfield  {journal} {\bibinfo  {journal}
  {arXiv preprint}\ } (\bibinfo {year} {2022})},\ \Eprint
  {https://arxiv.org/abs/2202.02773} {arXiv:2202.02773 [astro-ph.CO]}
  \BibitemShut {NoStop}%
\bibitem [{\citenamefont {Lewis}(2019)}]{Lewis:2019xzd}%
  \BibitemOpen
  \bibfield  {author} {\bibinfo {author} {\bibfnamefont {A.}~\bibnamefont
  {Lewis}},\ }\href@noop {} {\bibfield  {journal} {\bibinfo  {journal} {arXiv
  preprint}\ } (\bibinfo {year} {2019})},\ \Eprint
  {https://arxiv.org/abs/1910.13970} {arXiv:1910.13970 [astro-ph.IM]}
  \BibitemShut {NoStop}%
\end{thebibliography}%
\end{document}